\documentclass[utf8]{frontiersSCNS} 

\usepackage{url,hyperref,lineno,microtype,subcaption}
\usepackage[onehalfspacing]{setspace}

\def\keyFont{\fontsize{8}{11}\helveticabold }
\def\firstAuthorLast{Gibson {et~al.}} 
\def\Authors{Sarah E. Gibson\,$^{1,*}$, Angelos Vourlidas\,$^{2}$, Donald M. Hassler\,$^{3}$, Laurel A. Rachmeler\,$^{4}$, Michael J. Thompson\,$^{1}$, Jeffrey Newmark\,$^{5}$, Marco Velli\,$^{6}$, Alan Title\,$^{7}$, and Scott W. McIntosh\,$^{1}$}
\def\Address{$^{1}$High Altitude Observatory, NCAR, Boulder, CO, USA \\
$^{2}$Applied Research Laboratory, Johns Hopkins University, Laurel, MD, USA \\
$^{3}$Southwest Research Institute, Boulder, CO, USA \\
$^{4}$NASA Marshall Space Flight Center, Huntsville, AL, USA \\
$^{5}$NASA Goddard Space Flight Center, Greenbelt, MD, USA \\
$^{6|}$University of California, Los Angelos, CA, USA \\
$^{7}$Lockheed Martin Solar and Astrophysics Laboratory, Palo Alto, CA, USA.}
\def\corrAuthor{Sarah Gibson}

\def\corrEmail{sgibson@ucar.edu}

\begin{document}
\onecolumn
\firstpage{1}

\title[New Views on the Sun]{Solar Physics from Unconventional Viewpoints 
} 

\author[\firstAuthorLast ]{\Authors} 
\address{} 
\correspondance{} 

\extraAuth{}

\maketitle

\begin{abstract}


\section{}
We explore new opportunities for solar physics that could be realized by future missions providing sustained observations from vantage points away from the Sun-Earth line. These include observations from the far side of the Sun, at high latitudes including over the solar poles, or from near-quadrature angles relative to the Earth (e.g., the Sun-Earth L4 and L5 Lagrangian points). Such observations fill known holes in our scientific understanding of the three-dimensional, time-evolving Sun and heliosphere, and have the potential to open new frontiers through discoveries enabled by novel viewpoints.

\tiny
 \keyFont{ \section{Keywords:} Sun, Solar Interior, Photosphere, Chromosphere, Corona, Heliosphere, Space Weather} 
\end{abstract}

\section{Introduction}\label{sec:intro}

Observations from satellite missions have transformed the field of solar physics. High-resolution observations with near-continuous temporal coverage have greatly extended our capability for studying long-term and transient phenomena, and the opening of new regions of the solar spectrum has made detailed investigation of the solar atmosphere possible. However, to date most solar space-based missions have been restricted to an observational vantage in the vicinity of the Sun-Earth line (SEL), either in orbit around the Earth or from the Sun-Earth L1 Lagrangian point. As a result, observations from these satellites represent the same geometrical view of the Sun that is accessible from the Earth. 

Understanding the interior structure of the Sun and the full development of solar activity would benefit greatly from fully three-dimensional monitoring of the solar surface, atmosphere and heliosphere. On the one hand, simultaneous spacecraft observations from multiple vantage points allow studies of the deep interior structure of the sun via stereoscopic helioseismology; on the other, distributed observations reveal the complete evolution of activity complexes and enhance space-weather predictions dramatically. 

In this paper we will present the rich variety of science that is enabled by multi-vantage observations.  In section \ref{sec:review} we review past and present missions that have ventured away from the SEL. In section \ref{sec:newscience} we discuss the science enabled by extra-SEL vantages, including solar dynamo studies, solar atmospheric global connections, and solar wind evolution and transient interactions.  In section \ref{sec:spaceweather} we discuss the benefits of multi-vantage observations for space-weather monitoring, both for the Earth and other planets, and in section \ref{sec:discover} we consider the discovery space opened by new viewpoints.   In section \ref{sec:future} we identify the remaining key gaps in our heliospheric great observatory related to extra-SEL observations, and discuss how these may be filled through future opportunities--including both planned near-term missions and game-changing missions for the next decade and beyond.  Finally, in section \ref{sec:conclusions} we present our conclusions.

\begin{figure}[h!]
\begin{center}
\includegraphics[width=14cm]{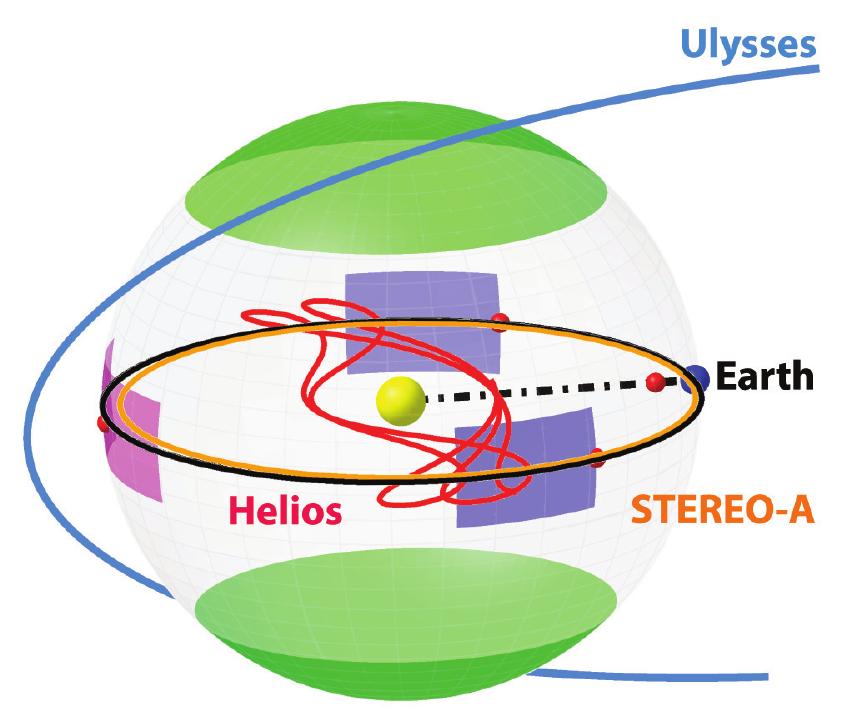}
\end{center}
\caption{An overview of interesting vantage points, highlighting previous extra-Sun-Earth-line (extra-SEL) missions. Sun and Earth are shown as yellow and blue dots, and Lagrange points are shown as red dots (radial scale deformed for clarity).  The SEL is indicated as the black dot-dashed line (intersecting the L1 Lagrange point and Earth).  The quadrature views are shown as blue patches (intersecting the L4 and L5 Lagrange points); the far-side view as a pink patch (intersecting the L3 Lagrange point); and the polar views as green patches. Sample STEREO orbit is shown in orange, Helios orbits in red, and Ulysses orbit in light blue.  The Helios  orbits are relative to the Earth, i.e., Heliocentric Earth Ecliptic (HEE) coordinates, the other orbits are in Heliocentric Inertial coordinates \citep{thompson_06}.
}\label{fig:oldorbits}
\end{figure}

\section{Extra-Sun-Earth-Line Observations to Date}\label{sec:review}

The first extra-SEL observations in the inner heliosphere were obtained by the two Helios spacecraft, launched in 1974 (Helios-1) and 1976 (Helios-2).  This mission obtained primarily in-situ measurements of solar wind plasma and cosmic rays \citep{gurnett_75,mcdonald_77,rosenbauer_77}, Faraday rotation measurements of light from external sources passing through the corona \citep[e.g.,][]{bird_85}, as well as remote sensing observations (but no imaging) of interplanetary dust \citep{leinert_81} and coronal mass ejections (CMEs) \citep{jackson_85}. The Helios spacecraft stayed close to the ecliptic plane; sample orbits are shown in Figure \ref{fig:oldorbits} (Helios-2 has the current record for proximity to the Sun at 0.29 AU).  Other extra-SEL, near-ecliptic in-situ solar wind measurements have been obtained by planetary missions, such as MESSENGER \citep{solomon_01}, Pioneer Venus Orbiter \citep{colin_80}, and the STEREO twin spacecraft \citep{kaiser_08}, which precede and follow the Earth’s orbit.

The corona and inner heliosphere (defined here at the space within 1 AU of the Sun) have been observed through extra-SEL, near-ecliptic imaging only since 2006 when the Sun-Earth Connection Coronal and Heliospheric Investigation (SECCHI) \citep{howard_08}, comprising an EUV imager (EUVI), two coronagraphs (COR1-2), and two heliospheric imagers (HI1, HI2), was deployed aboard both STEREO spacecraft (sample STEREO-A orbit is shown in Figure \ref{fig:oldorbits}).

\begin{figure}[ht!]
\begin{center}
\includegraphics[width=16cm]{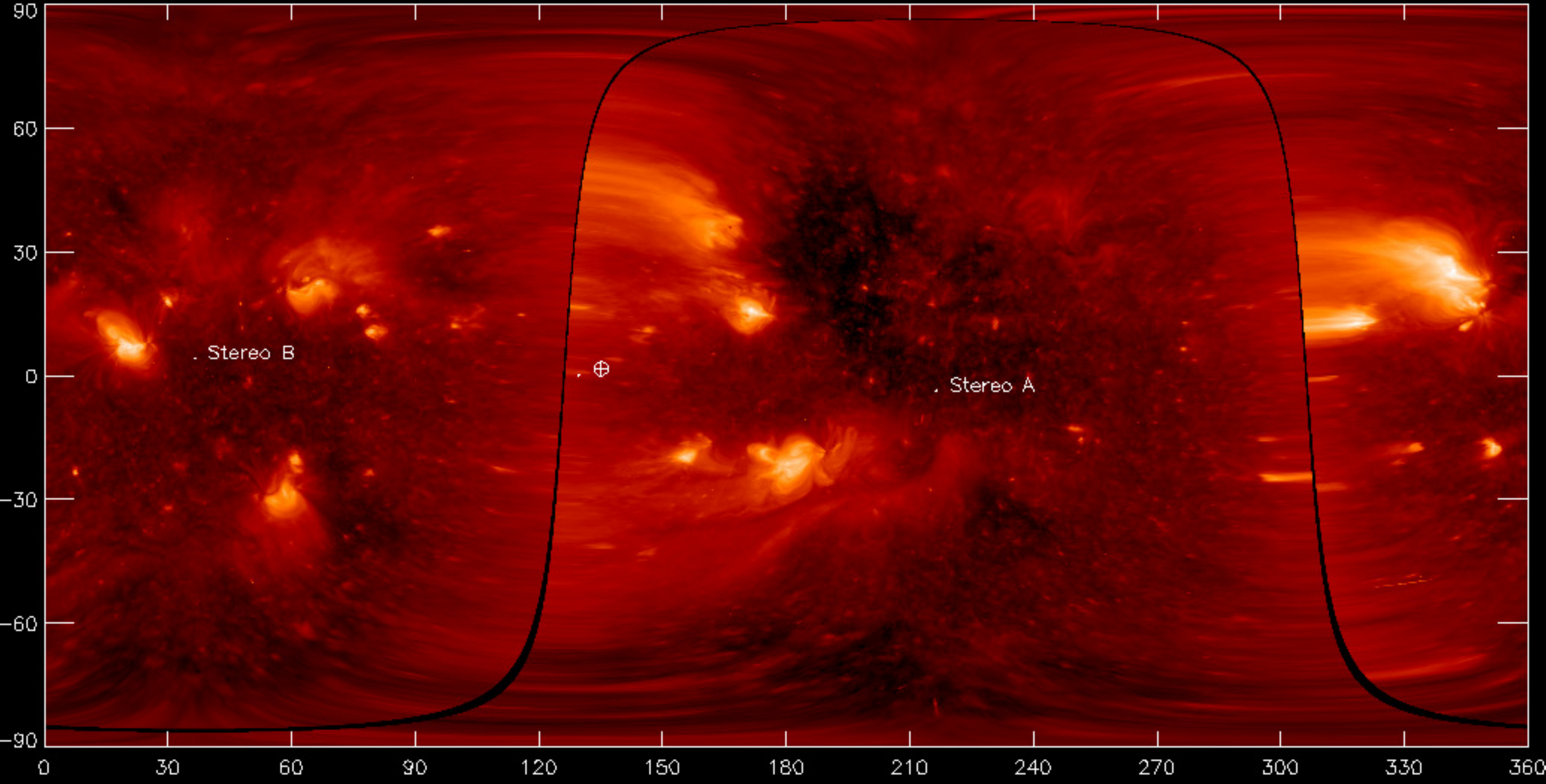}
\end{center}
\caption{A Carrington map of EUV synchronic observations of the full $360^\circ$ sun, made possible by far-side imaging provided by the twin STEREO spacecraft. Images were taken in the 284 \AA~channel on February 6, 2011 at 18:00 UT while the STEREO spacecraft were in exact opposition.
}\label{fig:synch}
\end{figure}

Since separating beyond $\approx 20^{\circ}$ from Earth in early 2008, STEREO began partial imaging of the far-side EUV corona. In 2011, the STEREO reached opposition and achieved the first ``synchronic'' observation of the full $360^{\circ}$ corona (Figure \ref{fig:synch}; see also Section \ref{subsec:corona}). Although STEREO did not carry any magnetographic capability, its $360^{\circ}$ coverage has enabled some degree of validation for flux transport models \citep{ugarte_15}  and helioseismic predictions of far-side flux emergence \citep{liewer_17}. As of 2018, STEREO-A continues to image the far-side and provide, with the support of SDO/AIA observations, instantaneous coverage of more that two-thirds of the corona.

Prior to STEREO, Earth-directed solar wind properties were generally derived indirectly by modeling and were validated by in-situ measurements.  The sparseness of inner heliospheric measurements considerably hindered the study of large-scale solar-wind structures, including stream-stream interaction regions--or corotating interaction regions (CIRs) in their evolved form -- and interactions among CMEs or between CMEs and the ambient solar wind. The continuously varying extra-SEL STEREO viewpoints have opened up research in many, previously elusive areas: from EUV coronal stereoscopy \citep{asch_11} and limb observations of the origins of Earth-directed CMEs \citep[e.g.,][]{patsour_16} to CIRs, CMEs entrained in CIRs \citep{rouillard_09}, and interacting CMEs \citep{sheeleyetal_08,lugaz_17}. The primary objective of the STEREO, driving its multi-viewpoint mission design, was to discover the 3D structure of CMEs. The power of this approach to recover CME sizes, directions, and detailed kinematics has been demonstrated by many studies \citep[e.g.,][]{thernisien_09,bosman_12,sachdeva_17}, to name a few). 

The only spacecraft to explore the heliosphere outside of the ecliptic plane was Ulysses  (1990-2009), which achieved a high-inclination ($80^{\circ}$) near-polar orbit with a Jupiter swing-by. Ulysses measured the solar wind plasma, electromagnetic fields, and composition in situ in the distance range between 1.3 and 5.4 AU from the Sun \citep{bame_92,balogh_92,gloeck_92}. Ulysses confirmed that the global, three-dimensional structure of the solar wind around solar minimum is made up of a fast, uniform flow in the northern and solar polar regions, and slow-to-medium speed wind streams originating in a coronal streamer belt that is confined to a relatively narrow range of latitudes on either side of the heliographic equator \citep{mccomas_98}. This band, in which the majority of solar wind variability is observed to occupy some $43^\circ$ in latitude, also contains the heliospheric current sheet (HCS) that separates the oppositely directed large-scale magnetic fields originating in the two hemispheres. During its second polar pass at solar maximum, Ulysses found a much more symmetric heliosphere with no systematic dependence on latitude \citep{mccomas_03}. The wind itself was generally slower and much more variable than at solar minimum at all latitudes. During its third and final polar pass, Ulysses found that the very strong solar minimum of 2008 did not lead to slower winds, but to a less dense wind (around 20\%) and lower momentum flux, while continuing to exhibit a generally dipolar structure \citep{issautier_08}. Ulysses also carried an X-ray/cosmic ray burst instrument (GRB) \citep{hurley_92} that was used in conjunction with measurements from the Hard-Xray telescope on the Yohkoh satellite \citep{kosugi_91} for stereoscopic analysis of solar flares \citep{kane_98}. Unfortunately Ulysses did not carry a Doppler magnetograph, so the details of solar polar flows and magnetic fields, and the weakening of the magnetic field over the last solar cycles, remain largely mysterious to this day.

\section{Science enabled by extra-Sun-Earth-Line Observations}\label{sec:newscience}

The scientific benefit of extra-SEL observations has been demonstrated to some extent by the missions described in Section \ref{sec:review}, but, in many cases, these observations merely whet our appetite for more comprehensive and sustained measurements.  We now describe the science enabled by extra-SEL observations, and identify areas where new observations are most desired.

\subsection{Solar dynamo: interior to surface}\label{subsec:interior}

While both the STEREO and Ulysses missions led to many scientific breakthroughs, there is no doubt that the lack of Doppler magnetographs on these extra-SEL spacecraft was a missed scientific opportunity. In addition to providing critical information about the surface magnetic field distribution and Doppler flow fields, these observations are the key inputs to helioseismic inversions. The power of helioseismology for probing the interior of the Sun has been proven by observations from views along the Sun-Earth line \citep{christdals_02,basu_08}, using both ground-based \citep[e.g.,][]{elsworth_94,harvey_96} and space-based \citep[e.g.][]{gabriel_97,kosovichev_97} observing platforms.

In particular, global helioseismology, which studies the Sun’s resonant modes manifesting as solar surface oscillations, revealed the Sun’s internal structure and rotation as never before.   The ``tachocline'', a boundary layer between the rigidly-rotating core and differentially-rotating convective envelope was deduced, and the variation of this differential rotation with latitude and depth probed \citep{thompson_03,christdals_07}.

By observing wave packets rather than resonant modes, local helioseismology provides another tool for imaging the time-varying structure and flows for local areas of the Sun \citep{gizon_10}.  For example, ring-diagram analysis measures distortions of the power spectrum of acoustic wave fields due to flows \citep{hill_88,schou_98}. Time-distance analysis \citep{duvall_93,kosovichev_96} determines sub-surface structure and flows by examining correlations between different points on the surface: note that both ends of a ray path must be resolved, so that single-vantage observations are necessarily limited to shallow waves and thus near-surface structures.  Finally, holography infers sub-surface inhomogeneities from perturbations of the acoustic wave field, and has been used to examine structure below sunspots as well as the emergence of active regions on the far side of the Sun \citep{lindsey_97,chang_97}. In contrast with global helioseismology, local helioseismology has the power to study how the solar structure and dynamics vary with longitude, and to distinguish differences between the northern and southern hemispheres.

In addition to helioseismology, observations of surface Doppler velocities and correlation tracking of magnetic features have been used to study meridional circulation, torsional oscillations, and supergranule patterns \citep{howe_09,rieutord_10}. These, along with observations of magnetic flux emergence over solar-cycle time scales provide inputs to surface-flux-transport and dynamo models \citep{mackay_12}.

Despite the strides made over the past two decades, important open questions about solar interior structure and flows remain. Helioseismology and photospheric measurements have provided a tantalizing glimpse of multi-cell flow patterns, with variation in both radius and latitude; however, different methods have given different results \citep{hanasoge_15}. In particular, because of the geometric limitations of the SEL vantage, there is uncertainty about the polar structure and flows, and about the strength and distribution of polar magnetic fields.

These represent critical observational constraints on solar dynamo models. They are particularly significant to the popular Babcock-Leighton/flux transport dynamo paradigm, as the cyclic reversals of polar fields in those models is sensitive to the structure of the high-latitude meridional flow, and to the magnetic flux budget associated with the transport and evolution of magnetic fields over the solar cycle \citep{charbonneau_10,hathaway_15}.  They also are relevant to understanding the fundamental physics of convection and its role in the dynamo. For example, observations from multiple vantages may reveal important differences in the structure and transport between polar and equatorial convective modes that determine global mean flow properties \citep{miesch_05} (see also discussion in Section \ref{sec:discover} regarding polar complexities ripe for discovery).  

Extra-SEL observations complement existing SEL observations in multiple ways (Table \ref{table:interior}). Multiple views increase surface coverage and help with disentangling different modes, improving signal-to-noise for both global and local helioseismic methods. They enable time-distance analysis of longer ray paths, probing deeper below the solar surface to resolve large-scale convection, giant cells, and possibly even the tachocline itself. Moving away from the SEL provides longitudinal coverage, yielding a more complete view of surface flow patterns and a comprehensive view of magnetic flux emergence and evolution.  Finally, observing from higher latitudes gives a direct view of polar fields and flows, and breaks north-south symmetries.
{\bf To make progress on the open questions on the solar dynamo, extra-SEL observations are needed to resolve solar surface-to-interior flows and magnetic fields as a function of longitude, latitude, and depth, and over the solar cycle.  The polar vantage is particularly helpful because of the critical information on polar fields and flows that it provides.}

\begin{table}[ht!]
\processtable{ }
{\begin{tabular}
{| m{2.5cm} || m{15.5cm} |}
  \hline
 \center{ {\bf Open Science Question} \vspace{.1cm} }  & \vspace{.1cm}  
What are the solar surface/interior flow and magnetic field patterns vs. longitude, latitude and depth, and how do they constrain dynamo models? \\ 
   \hline
    \center{ {\bf Measurements Needed}\vspace{.1cm}  } & \vspace{.1cm} 
1) Full-disk Doppler vector magnetographs (photosphere)
 \\ \hline
\center{{\bf Benefits from extra-SEL vantage} (assumes existence of complementary SEL observations)} &
 {\begin{tabular}
 {p{8cm} || p{1.5cm} || p{1.8cm} || p{1.8cm}}
& \vspace{.1cm} {\bf Polar} \vspace{.05cm}& \vspace{.1cm}{\bf Quadrature} \vspace{.05cm}& \vspace{.1cm}{\bf Far-side} \vspace{.05cm}
\\ \hline
\center{Helioseismic inversions able to probe deeper}& \vspace{.1cm} {\bf yes (1)} \vspace{.1cm}& \vspace{.1cm}{\bf yes (1)} \vspace{.1cm}& \vspace{.1cm}{\bf yes (1)} \vspace{.1cm}
\\ \hline
\center{Comprehensive view of flux emergence and evolution}& \vspace{.1cm} {\bf yes (1)} \vspace{.1cm}& \vspace{.1cm}{\bf yes (1)} \vspace{.1cm}& \vspace{.1cm}{\bf yes (1)} \vspace{.1cm}
\\ \hline
\center{
Polar fields and flows directly observed; North-South symmetry broken
}& \vspace{.1cm} {\bf yes (1)} \vspace{.1cm}& \vspace{.1cm} no \vspace{.1cm}& \vspace{.1cm} no \vspace{.1cm}
 \end{tabular}}{}
 \\ \hline
\end{tabular}}{}
\textbf{\refstepcounter{table}\label{table:interior} Table \arabic{table}.}{ Solar interior and dynamo studies enabled by extra-SEL observations.}
\end{table}

\subsection{Solar atmosphere: global connections
}\label{subsec:corona}

In addition to providing an upper boundary on the solar interior, Doppler magnetographs of the photosphere provide the lower boundary on the heliosphere.  Global models, ranging from potential-field-source-surface extrapolations \citep[e.g.,][] {wangsheeley_90,arge_03} to magnetohydrodynamic simulations \citep[e.g.,][]{toth_05,rileyetal_12} use photospheric observations to simulate the magnetic fields of the solar corona and heliosphere.

The single SEL vantage is a significant limitation, however.  Foreshortening affects both the poles and the East/West limbs. Magnetic fields on the far-side of the Sun are not seen at all, until they rotate into view. As a result, global magnetic boundary conditions on models cannot represent any single time, and indeed incorporate measurements taken over multiple days or even weeks. Standard synoptic maps are made up of an average in which each longitude is weighted towards observations taken when that longitude is at disk center \citep{hoeksema_86}. Alternatively, “synchronic” methods weight the entire map towards a single time, making use of flux-transport models and data assimilation \citep{worden_00,schrijver_03,arge_10,hickmann_15}.  Multiple views clearly are of great benefit to providing a synchronic view, as was shown in Figure \ref{fig:synch} using EUV STEREO observations. They are also important for obtaining a complete picture of magnetic flux emergence: this is critical for CME studies, since most occur in the 24 to 48 hours after new flux emerges. The addition of magnetograph measurements from a quadrature (L5) viewpoint is expected to significantly improve global magnetic simulations \citep{mackay_16}, as well as to improve modeling of the heliosphere \citep{pevtsov_16}.

Flux-transport models operating over solar-cycle time scales have also been used to model the evolution of polar magnetic fields.  It is difficult to validate such models, however.  Although the approximately seven-degree inclination of the Sun’s rotational axis with respect to the ecliptic plane means that observations from the Earth catch a glimpse of each solar pole once per year, they are still viewed from a very large angle, making accurate measurements of the polar photospheric magnetic fields extremely difficult \citep{petrie}. \citet{tsuneta_08} used high-resolution spectropolarimetric observations from the Solar Optical Telescope on board Hinode to resolve magnetic structures at the poles during a time of maximum polar inclination relative to the Earth, finding concentrated patches of strong field.  A comparison of the average magnetic flux implied by these high-resolution measurements to photospheric magnetic synoptic maps implies that the polar fields may be significantly underestimated, which could also explain inconsistencies with measurements of the open magnetic flux in the heliosphere \citep{linker_17}. 

In general, there remain substantial uncertainties about the photospheric magnetic boundary \citep{riley_14}. This affects not just global models but also magnetic models of particular coronal structures (e.g., active regions, prominences...) that are also sensitive to uncertainty in the magnetic boundaries \citep{Derosa09}. In addition to center-to-limb foreshortening issues, vector magnetic measurements from a single viewpoint possess intrinsic ambiguities \citep{semel_98}. These hamper our ability to determine the three-dimensional coronal magnetic field, and hence to investigate the roles of stored magnetic energy, magnetic helicity, and topology in solar eruptions. Coronal plasma and polarimetric measurements can be used to address such limitations by providing independent measurements for validation and/or optimization of coronal magnetic models \citep{Savcheva13b,malanushenko_14,gibson_16}.

The optically-thin corona is under-resolved by a single line of sight; multiple vantages enable more comprehensive tomographic and/or stereoscopic methods (see, e.g.,
 \citealp{Aschwanden2011} and references therein).  Such methods are useful for characterizing the morphology and spatial variation of the coronal density, temperature, and magnetic field (see, e.g., \citep{vasquez_11,asch_15,kramar_16b}).  Stereoscopy is also a powerful tool for distinguishing between models of flare acceleration and collisional and non-collisional transport processes \citep{casadei_17}  

\begin{figure}[h!]
\begin{center}
\includegraphics[width=11cm]{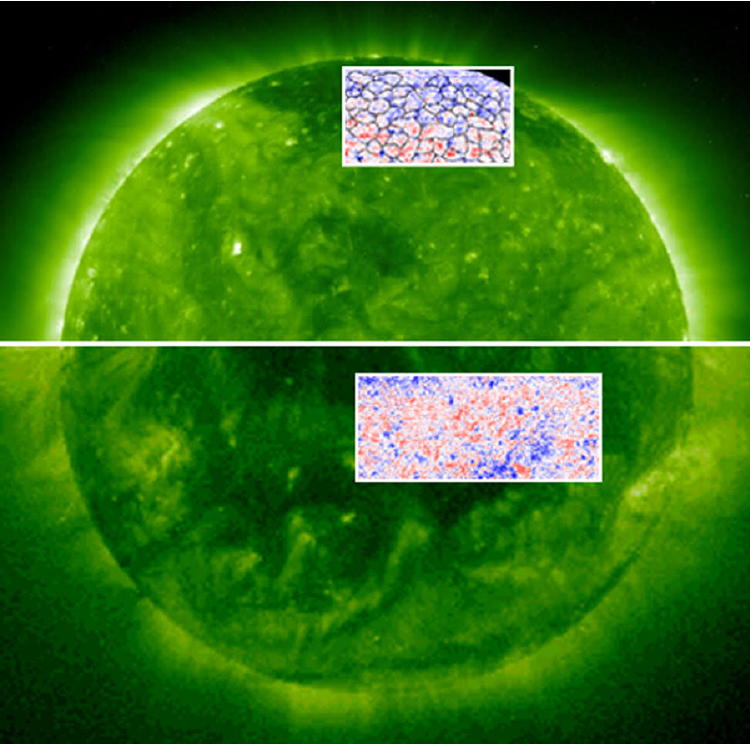}
\end{center}
\caption{Line-of-sight velocity structures observed by the SUMER instrument on SOHO show correlation of outflow velocity with chromospheric network structure within coronal holes both at mid-latitudes (bottom) and at high latitudes (top) \citep{hassler_99} As can be seen in the figure, observations of the detailed structure of solar wind outflow at the poles is made difficult by foreshortening, and would be significantly improved by out-of-the-ecliptic observations. 
}\label{fig:outflow}
\end{figure}

Modeling coronal structures observed at the limb is also fundamentally limited by the fact that observations of the underlying on-disk magnetic boundary cannot be co-temporal, whereas observations from the pole or quadrature enable simultaneous study of limb and disk.  This is particularly valuable for analyzing source regions of the solar wind using multiwavelength spectroscopy: both limb and disk observations are used to quantify the properties at the corona, and then the science is maximized via complementary in-situ solar-wind observations (including composition measurements). These include both solar wind transient structures and the ambient solar wind from coronal holes. Since EUV and UV spectrographs can only measure line-of-sight velocities, previous observations of the solar wind outflow in polar coronal holes with the SUMER spectrograph on SOHO \citep[Figure \ref{fig:outflow}]{hassler_99} have been limited at the poles. A polar mission could achieve direct measurements--both on-disk and in-situ--that explore fast solar wind sources and related structures, such as polar plumes associated with polar coronal holes \citep{deforest_97,hassler_97}, free from the effects of the foreshortening that has been inherent in observations to date \citep{mcintosh_06}.

Not only are there open questions about the global heliosphere and the localized coronal magnetic structures that are solar-wind source regions, but there are also questions about the nature of interactions between such local and global magnetic fields. One of the most tantalizing findings of multi-viewpoint imaging is the evidence for long-range interactions between eruptive events occurring over the span of days across a full solar hemisphere \citep{schrijver_11,Titov12,titov_17}. While the subject of ``sympathetic'' flaring and remote triggering of flares or eruptions has been discussed for decades in the literature, it can not be unambiguously resolved without synoptic $360^\circ$ coverage of the corona. Similarly, large-scale waves associated with eruptions often have global properties that require longitudinal coverage \citep[e.g.,][]{thompson_09,olmedo_12}.

Extra-SEL observations thus complement existing SEL observations in multiple ways (Table \ref{table:corona}). They improve coverage of the boundary of the heliosphere and of the birth-to-death evolution of solar structures, including global interactions.   More viewpoints also help vector magnetic field disambiguation and model optimization/validation, as well as stereoscopic/tomographic reconstructions of 3D coronal magnetic fields, thus yielding key information about magnetic energy, helicity, and topology. Finally, simultaneous observations of structures at the limb and of the boundary beneath them allows comprehensive studies of the coronal structures that are the source of the solar wind. {\bf To make progress on the open questions on the global connections of the solar atmosphere, extra-SEL observations are needed for better coverage and disambiguation of the vector magnetic boundary and for multi-vantage observations of optically-thin structures.  The polar vantage is particularly important to resolve uncertainty in the magnetic boundary condition and solar-wind source regions at high latitudes, with implications for the open magnetic flux in the heliosphere.}

\begin{table}[ht!]
\processtable{ }
{\begin{tabular}
{| m{2.5cm} || m{15.5cm} |}
  \hline
 \center{ {\bf Open Science Question} \vspace{.1cm} }  & \vspace{.1cm}  
What is the structure of the global coronal/heliospheric magnetic field? What are the source regions of the solar wind? How is magnetic energy stored/released in eruption, and what is role of helicity/topology?  How do local and global solar magnetic fields interact? 
 \\ 
   \hline
    \center{ {\bf Measurements Needed}\vspace{.1cm}  } & \vspace{.1cm} 
    {\begin{tabular}
    {|p{13cm}|}
(1) Full-disk Doppler vector magnetographs (photosphere)\\
(2) Chromospheric spectropolarimeters \\
(3) Full-Sun multiwavelength coronal  imagers\\
(4) Multiwavelength coronal spectrometers \\
(5) Polarimetric coronagraphs\\
(6) White-light/multiwavelength coronagraphs\\
(7) Heliographic imagers (HIs) with polarizers\\
(8) In-situ solar wind measurements
\end{tabular}}{}

 \\ \hline
\center{{\bf Benefits from extra-SEL vantage} (assumes existence of complementary SEL observations)} &
 {\begin{tabular}
 {p{8cm} || p{1.5cm} || p{1.8cm} || p{1.8cm}}
& \vspace{.1cm} {\bf Polar} \vspace{.05cm}& \vspace{.1cm}{\bf Quadrature} \vspace{.05cm}& \vspace{.1cm}{\bf Far-side} \vspace{.05cm}
\\ \hline
\center{Better coverage of global magnetic boundary; *Better constraints on contribution of polar fields to heliospheric open flux}& \vspace{.1cm} {\bf yes* (1; 8)} \vspace{.1cm}& \vspace{.1cm}{\bf yes (1)} \vspace{.1cm}& \vspace{.1cm}{\bf yes (1)} \vspace{.1cm}
\\ \hline
\center{Magnetic vector boundary $180^\circ$ disambiguation}& \vspace{.1cm} {\bf yes (1 - 2)} \vspace{.1cm}& \vspace{.1cm}{\bf yes (1 - 2)} \vspace{.1cm}& \vspace{.1cm} no \vspace{.1cm}
\\ \hline
\center{
More cotemporal lines of sight to reconstruct 3D coronal structures
}& \vspace{.1cm} {\bf yes (3 - 6)} \vspace{.1cm}& \vspace{.1cm} {\bf yes (3 - 6)} \vspace{.1cm}& \vspace{.1cm} no \vspace{.1cm}
\\ \hline
\center{
Boundary/limb view of structures observed simultaneously; solar-wind source regions connected to in-situ measurements; global interactions comprehensively observed
}& \vspace{.1cm} {\bf yes (1 - 6)} \vspace{.1cm}& \vspace{.1cm} {\bf yes (1 - 6)} \vspace{.1cm}& \vspace{.1cm} no \vspace{.1cm}
 \end{tabular}}{}
 \\ \hline
\end{tabular}}{}
\textbf{\refstepcounter{table}\label{table:corona} Table \arabic{table}.}{ Solar atmosphere/global connections studies enabled by extra-SEL observations.}
\end{table}

\begin{figure}[h!]
\begin{center}
\includegraphics[width=16cm]{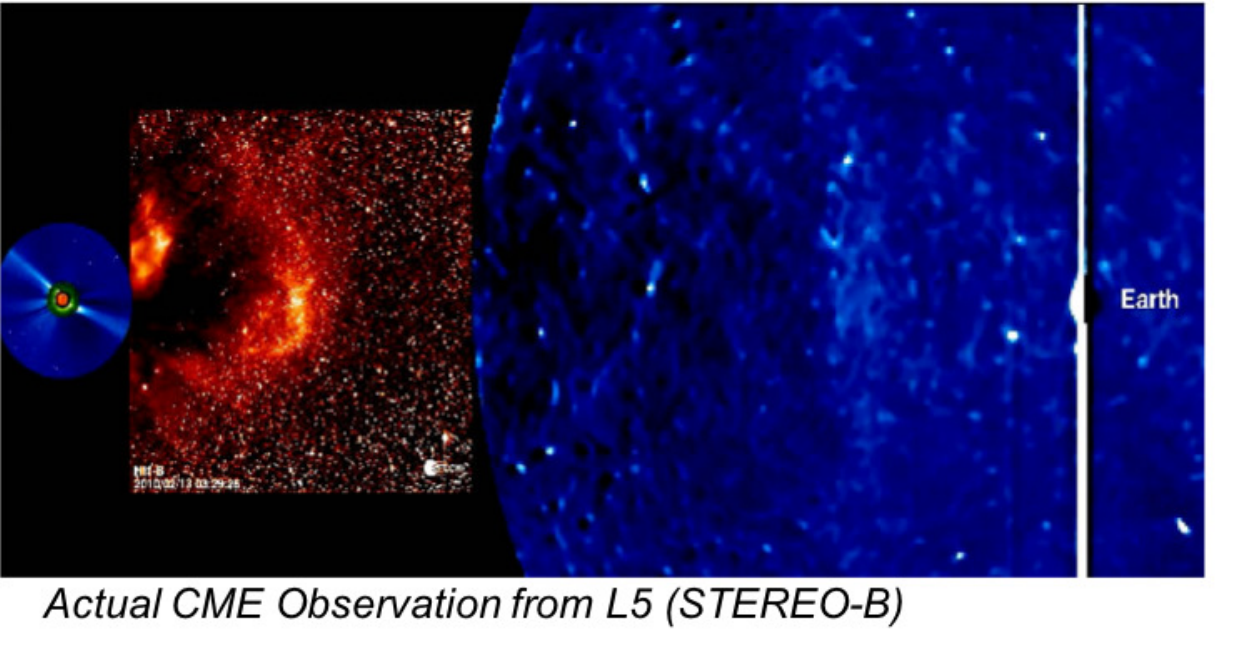}
\end{center}
\caption{Snapshot of an Earth-directed CME from the STEREO-Behind imagers while crossing the L5 point (02/13/2010 3:29 UT). The figure is a composite of the fields of view (FOVs) of all SECCHI imagers. The CME is in the HI1-B FOV at this time. The Earth is bright and saturates a few columns in the HI2-B FOV. The faint broad front of a preceding CME can be discerned in HI2-B.
}\label{fig:sunearth}
\end{figure}

\subsection{Solar wind and transients: evolution and interactions
}\label{subsec:helio}

As discussed in Section \ref{sec:review}, STEREO revolutionized heliospheric physics by providing consistent, spatially resolved imaging of the inner heliosphere with coverage from the Sun to 1 AU. When viewing from the extra-SEL, the STEREO heliospheric imagers can observe CIRs and CMEs en route to Earth without interruption (Figure \ref{fig:sunearth}), enabling studies of their evolution and interactions, and naturally feeding an increasingly sophisticated space-weather research field (see further discussion in Section \ref{sec:spaceweather}. 

While the STEREO mission demonstrates that we can indeed image the solar wind and its major components, it has left us with many unanswered questions. With regards to CMEs, heliospheric imaging analyses provide some indications of CME rotational evolution en route to Earth \citep[e.g.,][]{isavnin_14} and multiple cases of CME-CME interactions \citep[and references therein]{lugaz_17}, but it is hard to establish the reliability and general applicability of these results due to the long integration times and low spatial resolution of the current imagers on STEREO. Projection effects, long lines of sight through overlapping structures, and potentially complex internal morphology complicate the interpretation. Evolutionary effects such as ambient plasma pileup can also modify the brightness distribution around the transient, and skew triangulation or 3D reconstruction results.  This is particularly relevant to longitudinal deflections \citep{wang_04,isavnin_14}, as their study relies almost exclusively on imaging analysis. Extra-SEL/SEL coronagraph and heliospheric imagers are needed with better sensitivity, and with polarization capabilities to localize plasma in 3D \citep{deforest_16}. For maximum scientific return these need to be coupled with the complementary SEL/extra-SEL in-situ measurements during CME-CME (or equally, CME-CIR) interactions.  

Existing observations of the corona and inner heliosphere have also provided us with tantalizing clues that the laminar flow of the slow solar wind, revealed by the familiar streamer structure in coronagraph images, breaks into a more turbulent flow beyond 15-20 Rs \citep{deforest_14}. This is the height range where the wind may become super-Alfv\'enic, according to models \citep{verdini_09,zhaohoek_10,katsikas_10,goelzer_14,tasnim_16}. This is also the height, however, where current observations transition from a coronagraph to a heliospheric imager with lower spatial resolution and longer integration times. It is, therefore, unclear whether the change in the solar wind appearance is due to instrumental limitations or reflects a true change in the nature of the solar wind. In-situ measurements of the interplanetary magnetic field (IMF) in the corona, soon to be provided by the Parker Solar Probe \citep{fox_16} (Section \ref{subsec:planned}, will go a long way towards answering this issue. This will be done, however, over relatively limited spatial and temporal ranges. 

\begin{figure}[hb!]
\begin{center}
\includegraphics[width=15cm]{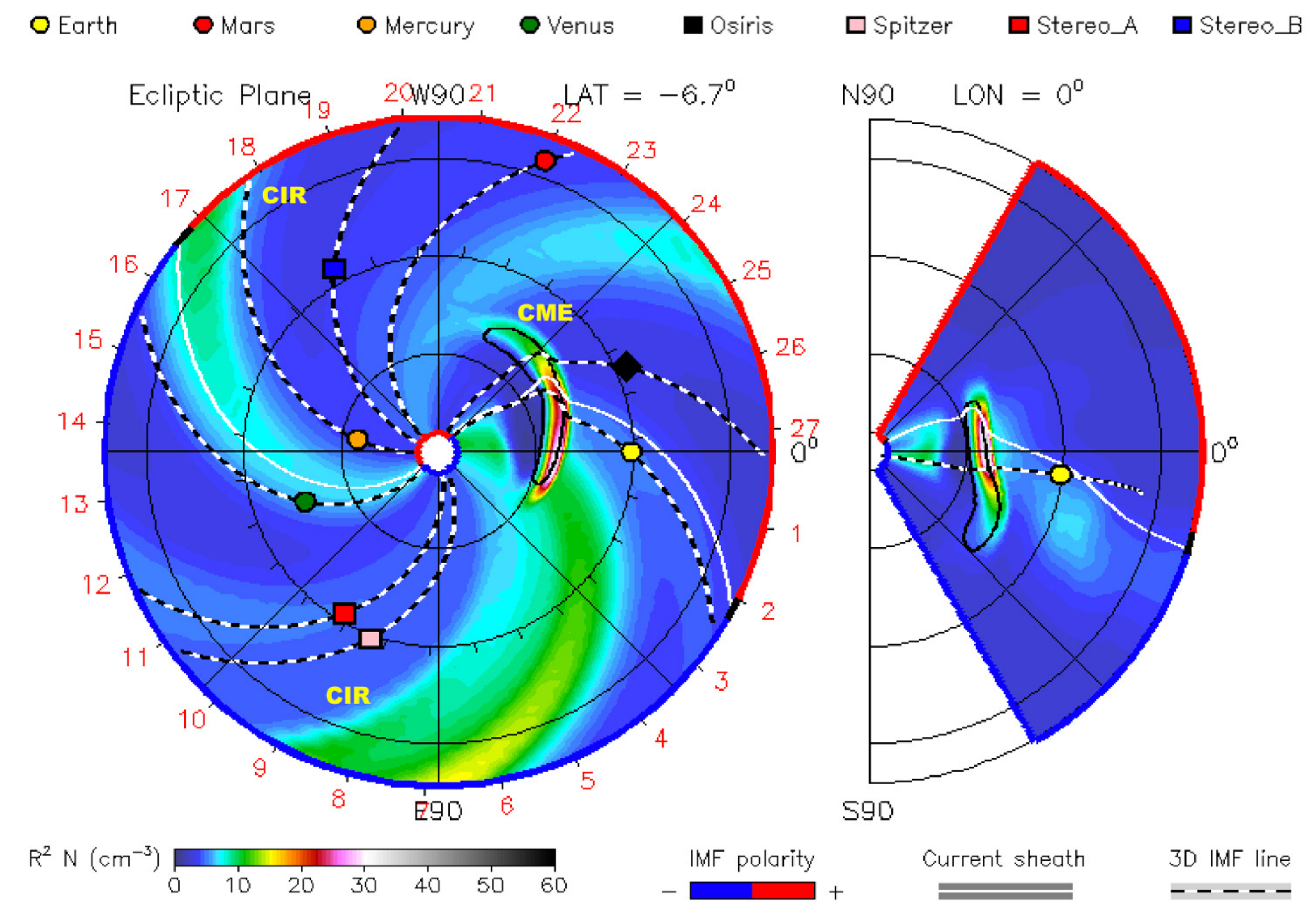}
\end{center}
\caption{Snapshot of the heliosphere, on 2/14/2018 00:00 UT,  from an ENLIL model run at the Community Coordinated Modeling Center. The normalized density is shown. Left panel: Polar view showing one CME and two CIRs. Earth is magnetically connected to the incoming CME. The interaction of the CME and a CIR is evident in this view. Right panel: Meridional cut at the location of Earth. It approximates the view from a heliospheric imager in Earth quadrature.
}\label{fig:enlil}
\end{figure}

Extra-SEL heliospheric observations complement existing SEL observations in multiple ways (Table \ref{table:helio}). Both  polar and quadrature viewpoints allow imaging of Earth-intersecting transient structures, and also permit in-situ measurements of structures imaged from the SEL.  In so doing they shed light on the nature of solar-wind/transient interactions, and enable important monitoring of space weather (see Section \ref{sec:spaceweather}). Capturing the large scale structure of the Alfv\'enic surface and its interplay with the outflowing solar plasma in a synoptic manner particularly benefits from the polar viewpoint. This viewpoint, unavailable even to the Solar Orbiter mission which will only reach $34^\circ$ above the ecliptic \citep{mueller_13} (Section \ref{subsec:planned}), comprehensively reveals both the radial and longitudinal evolution of the solar wind and transient structures propagating through it (Figure \ref{fig:enlil}). It is particularly good for capturing the formation and evolution of CIRs and shocks in a continuous fashion and for investigating the magnetic connectivity across large swaths of space. {\bf To make progress on the open questions on CME/CIR propagation, their interactions and the role and nature of the ambient solar wind, we need spatially resolved coverage of the inner heliosphere--both in-situ and (critically) imaging--at temporal scales matching the evolutionary timescales of these phenomena (tens of minutes to hours), and from multiple vantage points. The polar vantage is particularly beneficial because of the wide coverage and unique perspective it provides. }

\begin{table}[ht!]
\processtable{ }
{\begin{tabular}
{| m{2.5cm} || m{15.5cm} |}
  \hline
 \center{ {\bf Open Science Question} \vspace{.1cm} }  & \vspace{.1cm}  
How do transients evolve and interact with the ambient solar wind as they move through the heliosphere?
 \\ 
   \hline
    \center{ {\bf Measurements Needed}\vspace{.1cm}  } & \vspace{.1cm} 
    {\begin{tabular}
    {|p{13cm}|}
(6) White-light/multiwavelength coronagraphs\\
(7) Heliographic imagers (HIs) with polarizers\\
(8) In-situ solar wind measurements
\end{tabular}}{}
 \\ \hline
\center{{\bf Benefits from extra-SEL vantage} (assumes existence of complementary SEL observations)} &
 {\begin{tabular}
 {p{8cm} || p{1.5cm} || p{1.8cm} || p{1.8cm}}
& \vspace{.1cm} {\bf Polar} \vspace{.05cm}& \vspace{.1cm}{\bf Quadrature} \vspace{.05cm}& \vspace{.1cm}{\bf Far-side} \vspace{.05cm}
\\ \hline
\center{More lines of sight to reconstruct 3D solar-wind structures }& \vspace{.1cm} {\bf yes (6; 7)} \vspace{.1cm}& \vspace{.1cm}{\bf yes (6 - 7)} \vspace{.1cm}& \vspace{.1cm} no \vspace{.1cm}
\\ \hline
\center{
Complementary in-situ probing and remote imaging of solar wind/transient evolution
}& \vspace{.1cm} {\bf yes (7 - 8)} \vspace{.1cm}& \vspace{.1cm}{\bf yes (7 -8)} \vspace{.1cm}& \vspace{.1cm} no \vspace{.1cm}
\\ \hline
\center{
Longitudinal structure of Alfv\'en surface and of CMEs, CIRs, and shocks revealed
}& \vspace{.1cm} {\bf yes (6 - 7)} \vspace{.1cm}& \vspace{.1cm} no \vspace{.1cm}& \vspace{.1cm} no \vspace{.1cm}
 \end{tabular}}{}
 \\ \hline
\end{tabular}}{}
\textbf{\refstepcounter{table}\label{table:helio} Table \arabic{table}.}{ Solar-wind/heliospheric transient studies enabled by extra-SEL observations
}
\end{table}

\section{Space-weather significance of extra-Sun-Earth-Line observations
}\label{sec:spaceweather}

In addition to all that may be gained scientifically from extra-SEL observations, there are clear benefits to such measurements for space-weather prediction and monitoring.  

First of all, global coverage enables analysis of active longitudes \citep{detoma_00}, which have been connected to particularly strong and sustained space-weather activity \citep{castenmiller_86,bai_87,mcintosh_15}.  Similarly, longitudinal coverage is important for monitoring the development and evolution of long-lived low-latitude coronal holes, which can be the source of repeating high-speed solar wind streams that drive periodic geomagnetic, upper atmosphere, and radiation belt responses \citep{gibhss}. The benefits extend to irradiance measurements, which are limited near the solar limbs. Monitoring irradiance variations from a quadrature viewpoint would improve the short-term (7+ days) forecast accuracy of thermospheric and ionospheric density \citep{vourlidas_15}, important for satellite collision avoidance analyses and military applications. The current 7+ day forecasts lie outside the acceptable range of many users, especially around the maximum of solar activity \citep{vourlidas_18}.  The additional benefits of irradiance measurements from a polar viewing spacecraft would be less significant, since most irradiance variation occurs at active region latitudes, which are still close to the limb of a polar viewer. (However, see Section \ref{sec:discover} for discussion of discovery irradiance science from the polar vantage).

Beyond this, limb observations from the poles or quadrature provide a variety of information to improve space-weather forecasts. Before the eruption, these views enhance predictive capability for impending SEL-directed CMEs. For example, the presence of a tear-drop morphology in coronal cavities seen at the limb has been shown to indicate near-term eruption \citep{forland_13}. Once the eruption occurs, these views improve forecasts of CME geo-effectiveness. The so-called ``stealth'' CMEs \citep{robbrecht_09} have no obvious coronal sources when viewed from the SEL, but such Earth-directed CMEs would be readily detected from a coronagraph at a polar or quadrature vantage. These extra-SEL viewpoints also more accurately measure CME speed and mass (and hence, kinetic energy). For example, the accuracy of the CME time of arrival at Earth has improved to 6-7 hours from 24 hours in the pre-STEREO era \citep{colaninno_13}. There are early indications that the momentum flux of the Earth-directed part of a CME front, measured from a extra-SEL coronagraph, can be extrapolated to 1 AU and used to predict Dst variations \citep{savani_13}.

\begin{figure}[h!]
\begin{center}
\includegraphics[width=15cm]{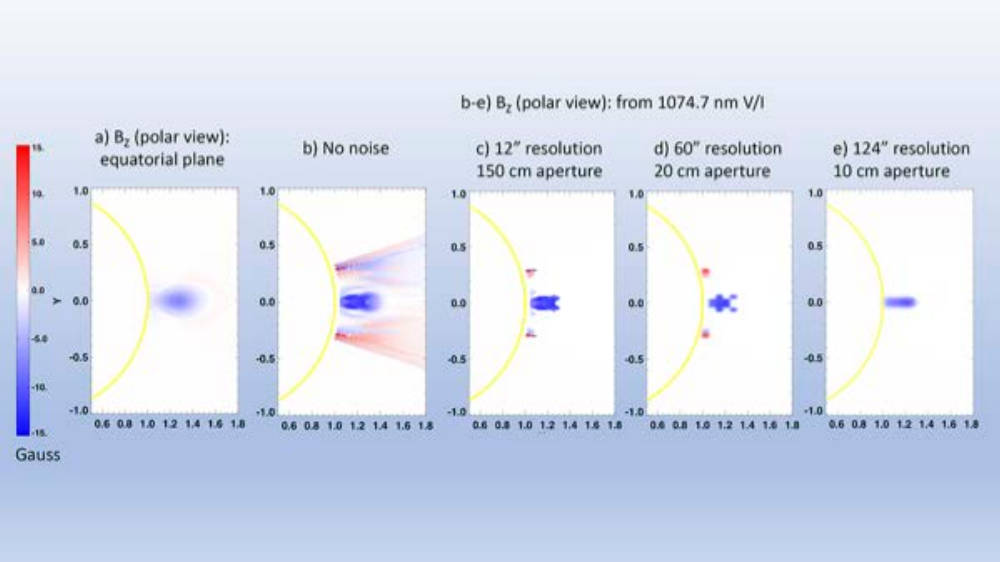}
\end{center}
\caption{The North-South component of a simulated, erupting CME \citep{fan_18} as viewed from the solar North pole. a) Simulation ground truth $B_z$ in equatorial plane. b) $B_z$ inverted from ratio of forward-modeled \citep{gibson_16} Stokes circular polarization V and intensity I. c) $B_z$ values with signal-to-noise ratio $> 3$, based on a 1.5 meter telescope,12” spatial resolution, and 5 minute integration. This is essentially the same as \citet{fan_18} Figure 12c, except without the background scatter included in that paper appropriate to ground-based telescopes. d) Same except for 20 cm telescope and 60” resolution. e) Same except for 10 cm telescope and 124” resolution. Note that the sign and strength of the pre-eruptive core field is captured for all.
}\label{fig:bzogram}
\end{figure}

Of utmost importance, extra-SEL configurations seem to be our best option for quantifying the magnetic field entrained in the CME. We note that from the poles, the line-of-sight field is $B_z$, the North-South component known to greatly impact geoeffectiveness. Line-of-sight $B_z$ could potentially be obtained, even mapped, via Faraday rotation measured from beacon signals sent from spacecraft distributed in and away from the ecliptic \citep[e.g.,][]{jensen_13}. Or, as recent analysis of a simulated CME has demonstrated, coronal IR spectropolarimetry has the potential to observe line-of-sight magnetic field strength at the core of an erupting CME (Figure \ref{fig:bzogram} \citep{fan_18}.

In addition to improved forecast capability, extra-SEL measurements are the best means for monitoring evolution of Earth-intersecting CMEs (and CIRs) via heliospheric imaging, as discussed in Section \ref{subsec:helio}. The STEREO HIs demonstrated that it is possible to spatially resolve and track CMEs to 1 AU \citep{davis_09,deforest_13}, but they did so at the expense of long exposure and low spatial resolution. Future measurements with improved sensitivity and polarization capability could resolve the inner structure of CMEs, measure the standoff distance between CMEs and their shocks, investigate in detail the interaction between CMEs and establish whether CMEs distort, rotate, and/or change propagation direction.  

For many of these measurements, the optimal viewpoint lies outside the ecliptic plane (see Table~\ref{table:spaceweather}. Imaging of the inner heliosphere from a polar viewing spacecraft along the solar rotation axis should provide direct evidence of the Parker spiral structure of the heliosphere, allow the precise mapping of CIRs and CMEs relative to them, and should enable study of the angular momentum coupling between the Sun and CMEs.  Moreover, as Figure \ref{fig:enlil} shows, full $360^\circ$ longitudinal coverage from a polar viewing spacecraft provides space-weather monitoring for all the planets and spacecraft in the inner heliosphere, not just the Earth-Moon system and L1. This benefit will become more and more important as human exploration expands out into the heliosphere. In particular, future human exploration to Mars and beyond will require heliosphere-wide space weather monitoring, forecasting, and early-warning outside of the SEL. A polar viewing spacecraft would be able to provide this capability.

Finally, as described in Section \ref{subsec:corona}, extra-SEL line observations of the photospheric and chromospheric magnetic field improves the boundary condition used in models of both localized source regions and the global solar wind through which transients propagate. Multi-viewpoint measurements of coronal properties can be used to further constrain and drive operational models, which can then be validated with in-situ measurements at the Earth. {\bf Paired with extra-SEL coronagraph/heliospheric imaging and tracking of the CME core, extra-SEL magnetic observations and data-constrained models could lead to a comprehensive system for forecasting CME impact \citep[e.g.,][]{savani_15,savani_17}. The quadrature vantage provides improved irradiance measurements enabling better modeling of the thermosphere/ionosphere. The polar vantage provides benefits to both Earth and planetary space-weather prediction and monitoring.}

\begin{table}[ht!]
\processtable{ }
{\begin{tabular}
{| m{2.5cm} || m{15.5cm} |}
  \hline
 \center{ {\bf Open Science Question} \vspace{.1cm} }  & \vspace{.1cm}  
How can we improve prediction of space weather on time scales of days, weeks, months, or longer? How can we improve forecasts of space-weather impacts at the Earth and throughout the solar system?
 \\ 
   \hline
    \center{ {\bf Measurements Needed}\vspace{.1cm}  } & \vspace{.1cm} 
    {\begin{tabular}
    {|p{13cm}|}
(1) Full-disk Doppler vector magnetographs (photosphere)\\
(2) Chromospheric spectropolarimeters \\
(3) Full-Sun multiwavelength coronal  imagers\\
(4) Multiwavelength coronal spectrometers \\
(5) Polarimetric coronagraphs\\
(6) White-light/multiwavelength coronagraphs\\
(7) Heliographic imagers (HIs) with polarizers\\
(8) In-situ solar wind measurements\\
(9) Faraday rotation (e.g., beacon signals between spacecraft and Earth)\\
(10) Irradiance monitors
\end{tabular}}{}

 \\ \hline
\center{{\bf Benefits from extra-SEL vantage} (assumes existence of complementary SEL observations)} &
 {\begin{tabular}
 {p{8cm} || p{1.5cm} || p{1.8cm} || p{1.8cm}}
& \vspace{.1cm} {\bf Polar} \vspace{.05cm}& \vspace{.1cm}{\bf Quadrature} \vspace{.05cm}& \vspace{.1cm}{\bf Far-side} \vspace{.05cm}
\\ \hline
\center{Longitudinal coverage enabling monitoring of developing coronal holes, active longitudes, etc. that drive “seasons” of space weather}& \vspace{.1cm} {\bf yes (1; 3; 6)} \vspace{.1cm}& \vspace{.1cm}{\bf yes (1; 3; 6)} \vspace{.1cm}& \vspace{.1cm}{\bf yes (1; 3; 6)} \vspace{.1cm}
\\ \hline
\center{
Improved modeling and monitoring of Earth-intersecting transients
}& \vspace{.1cm} {\bf yes (1 - 9)} \vspace{.1cm}& \vspace{.1cm}{\bf yes (1 - 9)} \vspace{.1cm}& \vspace{.1cm} no \vspace{.1cm}
\\ \hline
\center{
Improved modeling and monitoring of planet-intersecting transients
}& \vspace{.1cm} {\bf yes (1 - 9)} \vspace{.1cm}& \vspace{.1cm} {\bf sometimes (1 - 9)} \vspace{.1cm}& \vspace{.1cm} {\bf sometimes (1 - 9)} \vspace{.1cm}
\\ \hline
\center{
Line-of-sight measurements of southward-directed magnetic field
}& \vspace{.1cm} {\bf yes (5; 9)} \vspace{.1cm}& \vspace{.1cm} no \vspace{.1cm}& \vspace{.1cm} no \vspace{.1cm}
\\ \hline
\center{
Improved irradiance measurements
}& \vspace{.1cm}no \vspace{.1cm}& \vspace{.1cm} {\bf yes (3; 10)} \vspace{.1cm}& \vspace{.1cm} no \vspace{.1cm}
 \end{tabular}}{}
 \\ \hline
\end{tabular}}{}
\textbf{\refstepcounter{table}\label{table:spaceweather} Table \arabic{table}.}{ Space-weather prediction and modeling enabled by extra-SEL observations
}
\end{table}

\section{Discovery Space }\label{sec:discover}

Opportunities for discoveries by extra-SEL spacecraft abound. In addition to the broad range of discoveries likely to arise in association with the science topics described in the previous sections, previous experience shows that unique perspectives may be gained on celestial objects such as comets and asteroids: Ulysses, for example, crossed through the tail of one of the comets that was imaged--and detected in-situ--by STEREO \citep{fulle_07,neugebauer_07}. Extra-SEL viewpoints also can extend the observations of variable stars and improve the rate of discovery of exoplanets \citep{wraight_11}.

The polar view in particular offers unprecedented opportunities for discovery.  Stellar activity cycles are monitored and related to the Sun’s activity cycle through measurements, for example the Ca K line integrated over the solar disk as viewed from the Sun-Earth line \citep{wilson_78,baliunas_95,egeland_17}. How might these solar measurements change if obtained from the poles?  What are the implications for solar twins, which may be observed from a variety of angles relative to their rotation axes?  In general, observing from the poles could enable an unprecedented characterization of solar spectral irradiance, with as-yet-to-be-determined impact on our understanding of the Sun as a star, as well as on our understanding of stellar activity in general \citep{shapiro_14,shapiro_16}.
%

\begin{figure}[hb!]
\begin{center}
\includegraphics[width=11cm]{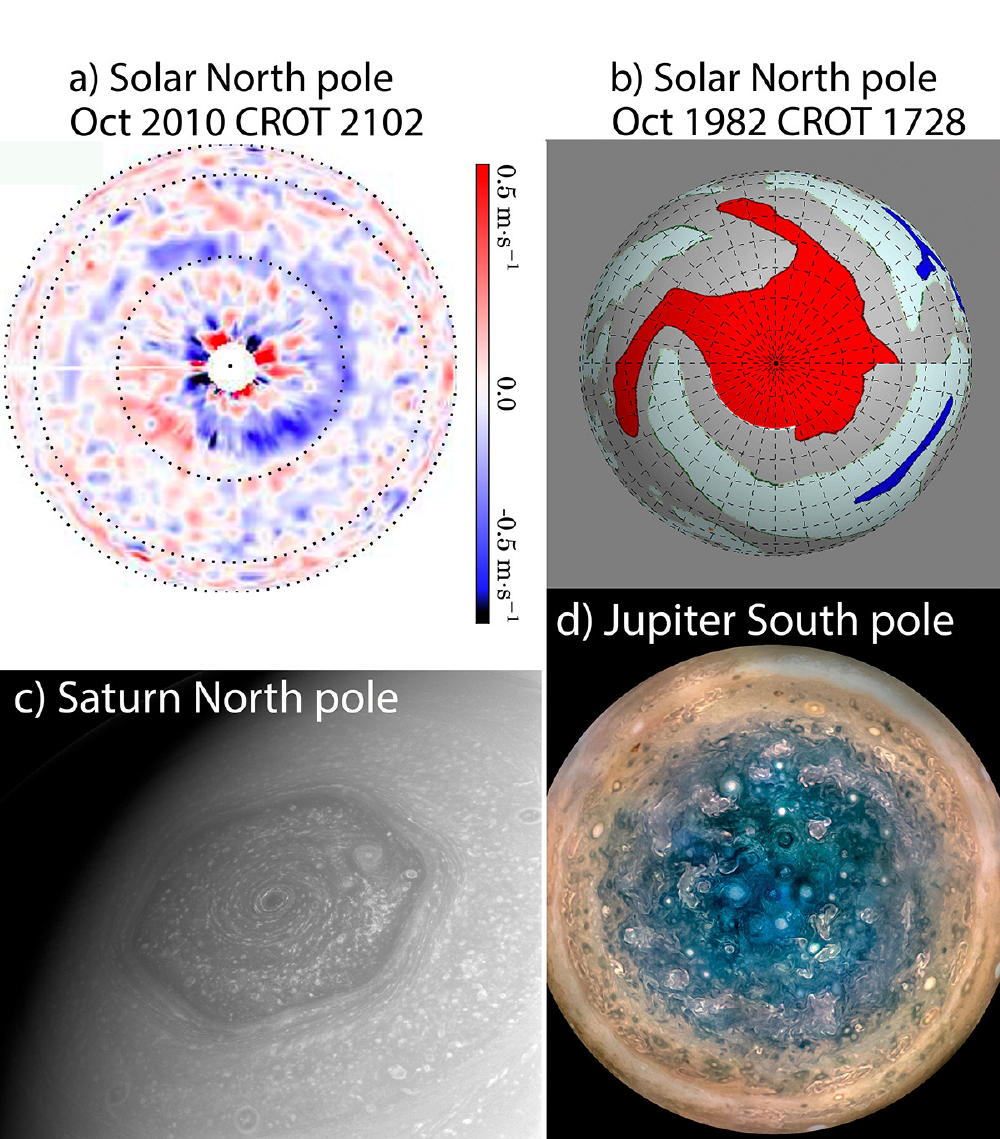}
\end{center}
\caption{(a) Ring-diagram analysis of near-surface flow anomalies indicate spiral flow patterns at the poles (from \citet{bogart_15}; Copyright \copyright{AAS}. Reproduced with permission). (b) Observations of large-scale, high-latitude magnetic features spanning multiple decades \citep{mcintosh_76,patmap} similarly demonstrate spiral structure at the poles (from \citet{webb_18}) (red=negative coronal hole, blue=positive coronal hole, grey = negative polarity quiet Sun, light blue = positive polarity quiet Sun, dark green=filaments).  (c) A view centered on Saturn's north pole. North is up and rotated $33^\circ$ to the left. The image was taken with the Cassini spacecraft wide-angle camera on June 14, 2013 using a spectral filter  sensitive to wavelengths of near-infrared light centered at 752 nanometers. (Source: NASA JPL) (d) Multiple images combined show Jupiter’s south pole, as seen by NASA’s Juno spacecraft from an altitude of 32,000 miles. The oval features are cyclones. (Credit: NASA/JPL - Caltech/SwRI/MSSS/Betsy Asher Hall/Gervasio Robles). 
}\label{fig:spiral}
\end{figure}

Beyond this--we have never observed the solar poles with imagers. What will we see? The semi-annual view of high latitudes arising from the tilt of the Earth’s orbit provides us with some sense of what we might expect in polar flow patterns, as does SEL observations of high-latitude large-scale magnetic features structures (Figure \ref{fig:spiral}A-B).  However, based on recent images from planetary missions, such as Juno and Cassini (Figure \ref{fig:spiral}C-D), direct observations of the poles is likely to reveal far more complex and beautiful structure than anything we have been able to piece together to date.

\begin{table}[ht!]
\processtable{ }
{\begin{tabular}
{| m{10cm} | m{4.4cm} |}
\hline
\center{{\bf Measurements needed} }& {\bf Vantages} \\ \hline
(1) Full-disk Doppler vector magnetographs (photosphere)

(2) Chromospheric spectropolarimeters 

(3) Full-Sun multiwavelength coronal  imagers

(4) Multiwavelength coronal spectrometers 

(5) Polarimetric coronagraphs

(6) White-light/multiwavelength coronagraphs

(7) Heliographic imagers (HIs) with polarizers

(8) In-situ solar wind measurements

(9) Faraday rotation (e.g., beacon signals between spacecraft and Earth)

(10) Irradiance monitors  
&
\center{Polar (P), 
Quadrature (Q),
Far-side (F)} 
\end{tabular}}{}
\end{table}
\begin{table}[h!]
\processtable{ }
{\begin{tabular}
{| m{5cm} | m{3.cm} | m{1.5cm} | m{2.cm} | m{1.5cm} |}
\hline
\center{{\bf Benefits from extra-SEL vantage} (assumes existence of complementary SEL observations)} &  \center{{\bf Measurements and vantages} (see above)} & \center{\bf{STEREO}} & \center{\bf{SO}} & {\bf PSP} \\ \hline
\center{Helioseismic inversions probe deeper} & \center{1 P/Q/F} & \center{no} & \center{{\bf limited}} & no \\ \hline
\center{Better coverage of global magnetic boundary; Comprehensive view of flux emergence and evolution} & \center{1 P/Q/F} & \center{no} & \center{{\bf limited}} & no \\ \hline
\center{Polar fields and flows directly observed; Magnetic vector boundary disambiguated} & \center{1 - 2 P;
1 -  2 P/Q} & \center{no} & \center{{\bf yes (1)}} & no \\ \hline
\center{Input to coronal 3D modeling} & \center{3 - 6 P/Q} & \center{{\bf yes (3; 6)}} & \center{{\bf yes (3; 4; 6)}} & no \\ \hline
\center{More lines of sight to reconstruct 3D solar wind structures} & \center{6 - 7 P/Q} & \center{{\bf yes (6 - 7')}} & \center{{\bf yes (6 - 7')}} & {\bf limited (7')} \\ \hline
\center{Simultaneous boundary/limb views of CMEs and precursors; global interactions comprehensively observed } & \center{1 - 6 P/Q} & \center{{\bf yes (3; 6)}} & \center{{\bf limited \\ (1; 3; 4; 6)}} &  no \\ \hline
\center{Solar-wind source regions connected to heliospheric imaging and in-situ measurements} & \center{1 - 8 P/Q} & \center{{\bf yes (3; 6; 7')}} & \center{{\bf yes (1; 3; 4; 6; 7'; 8)}} & {\bf yes (7'; 8)} \\ \hline
\center{Longitudinal structure of the Alfv\'en surface and of CMEs, CIRs, and shocks revealed} & \center{6; 7 P} & no & no & no \\ \hline
\center{
Improved monitoring and modeling of Earth-intersecting (planet-intersecting) transients.
} & \center{1 - 9 \\ P/Q \\(P/Q/F)
} & \bf{limited} (3; 6; 7’)& \bf{limited} (1; 3; 4; 6; 7'; 8) & \bf{limited} (7’; 8) \\ \hline
\center{
Longitudinal coverage enabling monitoring “seasons” of space weather
} & \center{1; 3; 6\\P/Q/F
} & \bf{limited} (3)& \bf{limited} (1; 3; 6) & no \\ \hline
\center{Improved irradiance measurements enabling better measurements of the thermosphere/ionosphere} & \center{3; 10 Q} & \bf{limited} (3) & no & no \\ \hline
\center{Line-of-sight measurement of southward-directed magnetic field} & \center{5; 9 P} & no & no & no \\ \hline
\end{tabular}}{}
\textbf{\refstepcounter{table}\label{table:gap} Table \arabic{table}.}{ Gap Analysis}
\end{table}

\section{Future missions and gap analysis}\label{sec:future}

We have shown that extra-SEL observations have potential for transformative progress in a range of science areas.  Tables \ref{table:interior} - \ref{table:spaceweather} describe the measurements needed from the various extra-SEL vantages that would lead to such progress.  We now discuss plans for future missions, and consider how the gaps in our current observational capability may be filled.

\subsection{Near-term  extra-SEL missions: Solar Orbiter and Parker Solar Probe
}\label{subsec:planned}

\begin{figure}[h!]
\begin{center}
\includegraphics[width=15cm]{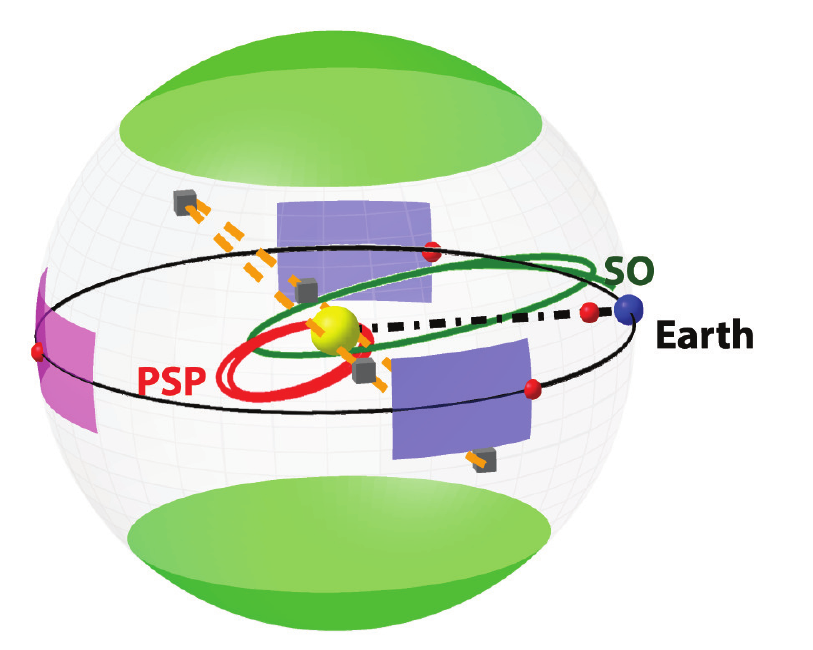}
\end{center}
\caption{Interesting viewpoints shown as in Figure \ref{fig:oldorbits}, with Parker Solar Probe (PSP; red) and Solar Orbiter (SO; green) orbits overlaid. The dashed orange line represents a sample “diamond” orbit with four spacecraft (grey cubes) $90^\circ$ apart, reaching as high as $75^\circ$ heliolatitude \citep{vourlidas_18}. The colored patches on the 1AU sphere are the same as in Figure \ref{fig:oldorbits}
}\label{fig:neworbits}
\end{figure}

The upcoming Parker Solar Probe (PSP), scheduled for launch in August of 2018, will explore the outer corona and inner heliosphere with very rapid solar encounters, starting at a perihelion of $35.7 R_{\odot}$  in November 2018 (already twice as close to the Sun as Helios ever went) and reaching its minimum perihelion of $9.86 R_{\odot}$ in late 2024 \citep{fox_16}.
In addition, the Solar Orbiter (SO) mission, to be launched no earlier than February 2020, will orbit the sun between $0.28-0.7 AU$ and reach a maximum inclination of  $\approx 30-34^\circ$ out of the ecliptic by the end of its extended mission in 2029-2030 \citep{mueller_13}. Figure \ref{fig:neworbits} shows sample orbits for both spacecraft.

Thanks to its gradually diminishing perihelion, PSP will directly probe the state of the solar wind as it evolves from its magnetically-dominated regime in the corona to the flow-dominated regime we experience at 1 AU. The critical heights where this occurs are collectively referred to as the ``Alfv\'en surface'', although density inhomogeneity may be so great as to disrupt any real smooth surface \citep{deforest18}. Theoretical models place the critical heights between $12-30 R_{\odot}$, depending on the location and phase of the solar cycle \citep{verdini_09,zhaohoek_10,katsikas_10,goelzer_14,tasnim_16}. These heights will become accessible to PSP within 2 years from launch and enable long-desired studies of the level of turbulence, heating, and kinematics of the primary constituents of the solar wind (i.e, electrons, protons, and helium). These, in turn, will unveil the energy flow from the sun to the heliosphere and help us pinpoint the mechanism responsible for solar wind heating and acceleration. PSP will spend 400 hrs below $15 R_{\odot}$ (75 hrs below  $12 R_{\odot}$) over 5 years and thus measure the trans-Alfv\'enic solar wind properties during a considerable part of the solar cycle and over a multitude of locations and activity levels. The radial and longitudinal scans, enabled by the close perihelia, will help to uncover the role of reconnection jets and instabilities in the heating and acceleration of the solar wind. 

In addition, PSP will provide the first measurements of the energetic particle populations in the solar corona and allow the separation of transport versus production effects in energetic particles. This has been a stumbling block in deciphering the mechanisms responsible for particle acceleration in astrophysical plasmas for decades. 

In close synergy, SO will obtain similar, but outside-the-ecliptic, in-situ measurements that will help to unravel the latitudinal structure and flow of energy (and energetic particles) in the inner heliosphere. Combined with the first ever direct measurements of the polar magnetic fields, the SO mission will considerably narrow uncertainty for global solar and heliospheric plasma and magnetic field modeling, leading to breakthroughs in many research areas, including solar dynamo evolution, energetic particle transport, and space-weather forecasting. Unlike PSP, SO has a full complement of solar and coronal imaging instruments capable of very high spatial resolution measurements ($< 300 km$) during perihelia to help uncover the physical processes behind eruptive events (i.e., magnetic reconnection) that are thought to occur at small spatial scales.  SO instruments also span a broad spectral range, and include an X-ray spectrometer that in conjunction with SEL observations will enable spectroscopic analysis of solar flares at high energy, helping to distinguish between flare acceleration models \citep{casadei_17}.  Finally, SO's strongest contributions will likely be to deciphering the connectivity between solar and heliospheric magnetic structures and plasmas, thanks to its unique combination of remote sensing and in-situ instruments.

Although we expect PSP and SO to lead to groundbreaking advances in several of the open science questions discussed above, the two missions will not fully address the open questions listed above for several reasons. PSP and SO are designed to meet specific science objectives--pertaining to the acceleration, heating, and structure of the solar wind--via {\it encounter mission} designs. This means that the instruments operate for limited amounts of time per orbit (10 days for PSP, 30 days for the SO remote sensing payload). Other complications include restricted data volumes due to the limited number of downlinks and onboard storage, data latency on the order of months, and angular configurations between the two spacecraft and Earth that are continuously changing. 

These operational constraints reduce the length of polar magnetic field observations by SO, for example, to only a few days per orbit (up to ten days per orbit by the end of the extended mission).. The shortness of this time period will limit helioseismology studies \citep{loeptien_15}. The data volumes and short observing periods also do not allow for synoptic observations or consistent $360^\circ$ coverage of the solar atmosphere from the SO imagers. PSP imaging is restricted to a wide field heliospheric imager providing context for the in-situ payload but has no disk imaging (for thermal reasons). 

It is undeniable that both missions will provide unique data and views of the coronal and heliospheric environment, and enhance the sophistication of multi-viewpoint analysis, far beyond STEREO. However, they are necessarily limited to the science achievable by the instruments they have on board, and in their ability to obtain the sustained measurements needed for longer-time-frame studies and space-weather monitoring (see Table \ref{table:spaceweather}). We now consider how next-generation extra-SEL mission concepts might address the remaining gaps between the observational capabilities of our existing and planned missions, and the outstanding science questions raised in this paper.

\subsection{Missions for the Future}\label{subsec:gap}

Several white papers describing extra-SEL mission concepts were submitted during the last Solar and Space Physics Decadal Survey activies, and were also summarized in the 2014 Heliophysics Roadmap. Variations of those concepts were submitted for the Next Generation Solar Physics Mission call for ideas. The majority of these white papers are not in the published literature, but a subset, with emphasis on helioseismology science, is discussed by \citep{sekii_15}. Here, we present some representative concepts for mission architectures that emphasize sustained measurements, and so fill most, if not all, of the gaps indicated by Table \ref{table:gap} (depending on instrument payload).

\subsubsection{Quadrature mission}
The idea of a mission to the Lagrange L5 point has been explored in several concepts in recent years \citep{webb_10b,gopal_11,vourlidas_15,lavraud_16}. It has obvious advantages for space-weather research and forecasting, such as more accurate speed measurements for Earth-directed CMEs and increased coverage of solar irradiance and the photospheric magnetic boundary for operational models. It also addresses many of the open science questions that particularly benefit from sustained measurements, such as probing the solar interior more deeply, and observing the evolution of structures over time, as well as 3D modeling of CMEs and their source regions.  Such analyses depend upon the existence of complementary SEL observations. 

From the mission design perspective, injection towards the L5 (and L4) points is relatively straightforward but requires significant $\Delta V$ and hence a large rocket. There is considerable trade space: orbit around L5 vs. stationed at L5, drift vs. direct injection, and travel time vs. mission length, to name a few. Indeed, it is also possible to put something at $90-120^\circ$ off the SEL utilizing a launch into a geo-transfer orbit and electric propulsion.

An L5 concept was studied for the Solar and Space Physics Decadal Survey, and estimated to cost above $600M (FY14$) for a standard spacecraft with multiple instruments. An operational space-weather mission to L5 could be cheaper if it had a reduced payload--at minimum, a coronagraph and a magnetograph--as was the case studied by \citet{trichas_15}. As of the writing of this paper, the European Space Agency (ESA) is in Phase-A development for an operational L5 mission. 

An innovative approach based on a fractionated spacecraft concept was also proposed by \citep{liewer_09}. Instead of a monolithic spacecraft carrying a multitude of instruments, the authors proposed the launch of a set of cubesats or minisats each carrying a single telescope along with a minisat carrying a standard antenna to relay communications from the constellation to Earth. The advantages include 1) lower launch costs through extensive use of hosted payload opportunities, 2) measurement persistence, since failed cubesats or telescopes could be replaced with another launch, and 3) redundancy, since a single spacecraft failure would not take down the whole constellation. The disadvantages include 1) the necessarily reduced size of instrument payloads, leading to restrictions on aperture size and other performance metrics, 2) data acquisition limitations -- since cubesats have less powerful radios and smaller antennas, and 3) the low Technology Readiness Levels (TRLs) for intra-spacecraft communications and of interplanetary cubesats. 

In summary, a mission to L5 or similar quadrature viewpoint seems to be the obvious next step after STEREO.  The benefits of this vantage for an operational space-weather mission are particularly strong.  The benefits for basic research are also high; however, as we will now discuss, a polar vantage achieves these same benefits, and more. 

\subsubsection{Solar Polar Imager (Single and constellation)}

Imaging of the solar poles has been a long-held desire of the solar community ever since the cancellation of the imaging sister payload to Ulysses. Mission concepts have been proposed for the ESA large mission competition \citep[e.g., POLARIS][]{appourchaux_09,appourchaux_14}, and the Chinese space program \citep[SPORT][]{xiong_14}.
The primary scientific objective of these missions is the unique opportunities they provide for measuring the evolving polar magnetic field and for investigating sub-surface flows beneath the poles via helioseismology. As Table \ref{table:gap} demonstrates, further unique science is achieved from a polar vantage pertaining to the longitudinal structure of the coronal streamer belt, Alfv\'en surface, and transient/solar wind interactions. Moreover, the space-weather and basic-research open questions pertaining to 3D modeling of CMEs and their source regions and solar wind interactions are also achieved from a polar vantage.  Finally, the polar vantage provides the most opportunities for discovery as discussed in Section \ref{sec:discover}.

Achieving high inclinations has traditionally been the stumbling block for the realization of such polar missions. A Jupiter gravity assist, as was done for the Ulysses mission, remains the most practical method for going over the poles. However, this approach also implies that aphelion occurs near Jupiter, resulting in a very long orbit with only a small fraction of time spent close to the Sun. The SPORT approach adds Venus-Jupiter gravity assists to reduce the perihelion to about 0.7 AU and shorten the orbit somewhat. However, time above the poles (defined here as above $60^\circ$) remains a small portion of the overall orbit. The window of science operations can be extended for orbits within 0.8 AU but the Venus gravity alone limits the inclination maximum to about $34^\circ$, with chemical propulsion. For this reason, many solar-polar mission concepts use solar sails (150m x 150m) to achieve high inclinations, presenting challenges associated with the large size and low TRL of solar-sail technology. 

Given the high scientific value and significant opportunity for discovery of a solar polar imaging mission, new mission concepts continue to be developed \citep{hassler_18a,hassler_18b}. In particular, advances in new spacecraft technology and instrument miniaturization are enabling new mission concepts to achieve the core science objectives of a polar imaging mission.  For example, a new version of a solar polar mission (Solar Polar Explorer (SOLPEX)) has been developed \citep{newmark_18} using a combination of a Jupiter gravity assist and solar electric propulsion (e.g., NASA Evolutionary Xenon Thruster (NEXT) thrusters) to achieve a high inclination ($> 75^\circ$) orbit with an eventual $\approx 1$ year period (final orbit is 0.5 x 1.5 AU). SOLPEX would contain a suite of remote sensing (Doppler magnetograph, EUV imager, coronagraph, and heliospheric imager) and in-situ instruments, and could complete three passes over each of the solar poles within a nine year mission lifetime, with each pass lasting $>38$ days above $60^\circ$.  

A wide range of instrumentation could reveal new information from a polar view (Table \ref{table:gap}). The fractionated approach might lend itself well to a polar strategy, with minisats carrying smaller instruments (e.g., magnetograph, EUV imager, and/or compact coronagraph) that would provide continuous $360^\circ$ coverage.  Larger missions could carry other, more state-of-the-art instrumentation to fill in the other science gaps that we have identified.  Such an approach would allow the participation of multiple international space agencies, so that over time, a truly global perspective on the Sun and heliosphere would be achieved.

A final consideration: a single spacecraft in polar orbit would still spend a finite time over the poles. Truly sustained polar observations could be achieved by the deployment of a constellation of spacecraft. For example, four spacecraft spaced $90^\circ$ apart in diamond formation (Figure \ref{fig:neworbits} and \citet{vourlidas_18}) would provide continuous, complete $360^\circ$ coverage of the solar atmosphere, from the photosphere to the corona and beyond (depending on the payload). Such a concept would also reduce the need for coverage in the ecliptic plane.

\section{Conclusions}\label{sec:conclusions}

Extra-SEL vantages present unique opportunities for answering the outstanding science questions of heliophysics, for improving space-weather monitoring and prediction, and for revealing new discoveries about our Sun and solar system.  A broad range of observations take on new significance away from the SEL, including both remote-sensing and in-situ measurements.  We have discussed the benefits of measurements from several types of instrumentation in this paper (Table~\ref{table:gap}), but have not attempted to put these in priority order.  A promising path forward in establishing the relative importance of different measurements is to combine comprehensive models of the solar physical state with forward modeling of observables (see, e.g., \citealp{gibson_15b}). In particular, sensitivity analyses that evaluate how different types of observations at different vantages constrain and improve model-data fitting are essential for establishing the most-urgently needed new measurements (see, e.g., \citealp{kramar_06,mackay_16,pevtsov_16}).

All three of the vantages considered in this paper--polar, quadrature, and far-side--are of interest. It should be noted, however, that Tables \ref{table:interior} - \ref{table:gap} considered these vantages independently, while a mission such as STEREO obtains both quadrature and and far-side vantages, and similarly, a mission in a high-latitude orbit may view the far side or quadrature as it crosses the ecliptic. Moreover, in Tables \ref{table:interior} - \ref{table:gap} we assumed only the availability of complementary SEL observations; clearly, if both quadrature and far-side measurements were available, many of the two-view benefits ascribed only to the quadrature-SEL or polar-SEL combinations would be realized.

Bearing this in mind, it is nevertheless a worthwhile exercise to consider the science enabled by the three vantages independently as we have done, and in particular to focus on the benefits of {\it sustained} observations from these vantages. The far-side and quadrature vantages have the advantage of being relatively easy to access, and the use of a Doppler magnetograph from both would enable global magnetic coverage and deeper helioseismic inversions. The quadrature vantage would also provide a second line of sight on structures visible along the SEL, enabling 3D reconstructions. A sustained view would allow constant monitoring of transients directed along the SEL, providing substantial space-weather operational benefit. The polar vantage is the most compelling overall, as it would achieve essentially the same capabilities as from quadrature, and also reveal the poorly observed poles as never before. For this reason, it is essential that investment in and demonstration of the technologies needed for polar missions are prioritized over the next few years.  In this way, the groundwork may be laid so that by the next decadal survey, community proposals for solar polar missions will not only be scientifically justified, but technologically robust.

\section*{Conflict of Interest Statement}

The authors declare that the research was conducted in the absence of any commercial or financial relationships that could be construed as a potential conflict of interest.

\section*{Author Contributions}

MJT, SEG, LAR, MV, AT, and SWM contributed to the initial conception and design of the paper. SEG wrote the main draft and coordinated inputs; AV, DMH, LAR, MJT, MV, and JN contributed text and figures. All of the authors read and critically revised the paper, approved the final version, and agreed to be accountable for all aspects of the work.

\section*{Funding}
NCAR is supported by the National Science Foundation.
A.V. is supported by NNX16AH70G under the STEREO MO\&DA and NRL grant N00173-16-1-G029.

\section*{Acknowledgments}
We thank Yuhong Fan for providing the simulation data cube used in Fig. \ref{fig:bzogram}, and Mark Miesch for many helpful discussions. We also thank Ricky Egeland for internal HAO review of the manuscript.

\bibliographystyle{frontiersinSCNS}

\begin{thebibliography}{142}
\providecommand{\natexlab}[1]{#1}
\expandafter\ifx\csname urlstyle\endcsname\relax
  \providecommand{\doi}[1]{doi:\discretionary{}{}{}#1}\else
  \providecommand{\doi}{doi:\discretionary{}{}{}\begingroup
  \urlstyle{rm}\Url}\fi
\providecommand{\selectlanguage}[1]{\relax}

\bibitem[{\textbf{{Appourchaux} et~al.}(2014)\textbf{{Appourchaux},
  {Auch{\`e}re}, and Antonucci}}]{appourchaux_14}
{Appourchaux}, T., {Auch{\`e}re}, F., and Antonucci, E.~e. (2014), in
  M.~{MacDonald}, ed., Advances in Solar Sailing, volume 259 (Springer: Berlin,
  Heidelberg), volume 259

\bibitem[{\textbf{{Appourchaux} et~al.}(2009)\textbf{{Appourchaux}, {Liewer},
  {Watt}, {Alexander}, {Andretta}, {Auch{\`e}re} et~al.}}]{appourchaux_09}
{Appourchaux}, T., {Liewer}, P., {Watt}, M., {Alexander}, D., {Andretta}, V.,
  {Auch{\`e}re}, F., et~al. (2009), {POLAR investigation of the Sun - POLARIS},
  \emph{Experimental Astronomy}, 23, 1079--1117,
  \doi{10.1007/s10686-008-9107-8}

\bibitem[{\textbf{{Arge} et~al.}(2003)\textbf{{Arge}, {Harvey}, {Hudson}, and
  {Kahler}}}]{arge_03}
{Arge}, C.~N., {Harvey}, K.~L., {Hudson}, H.~S., and {Kahler}, S.~W. (2003),
  {Narrow coronal holes in Yohkoh soft X-ray images and the slow solar wind},
  in M.~{Velli}, R.~{Bruno}, F.~{Malara}, and B.~{Bucci}, eds., Solar Wind Ten,
  volume 679 of \emph{American Institute of Physics Conference Series}, volume
  679 of \emph{American Institute of Physics Conference Series}, 202--205,
  \doi{10.1063/1.1618577}

\bibitem[{\textbf{{Arge} et~al.}(2010)\textbf{{Arge}, {Henney}, {Koller},
  {Compeau}, {Young}, {MacKenzie} et~al.}}]{arge_10}
{Arge}, C.~N., {Henney}, C.~J., {Koller}, J., {Compeau}, C.~R., {Young}, S.,
  {MacKenzie}, D., et~al. (2010), {Air Force Data Assimilative Photospheric
  Flux Transport (ADAPT) Model}, \emph{Twelfth International Solar Wind
  Conference}, 1216, 343--346, \doi{10.1063/1.3395870}

\bibitem[{\textbf{{Aschwanden}}(2011)}]{asch_11}
{Aschwanden}, M.~J. (2011), {Solar Stereoscopy and Tomography}, \emph{Living
  Reviews in Solar Physics}, 8, 5, \doi{10.12942/lrsp-2011-5}

\bibitem[{\textbf{Aschwanden}(2011)}]{Aschwanden2011}
Aschwanden, M.~J. (2011), Solar stereoscopy and tomography, \emph{Living
  Reviews in Solar Physics}, 8, 1, 5, \doi{10.12942/lrsp-2011-5}

\bibitem[{\textbf{{Aschwanden} et~al.}(2015)\textbf{{Aschwanden}, {Schrijver},
  and {Malanushenko}}}]{asch_15}
{Aschwanden}, M.~J., {Schrijver}, C.~J., and {Malanushenko}, A. (2015), {Blind
  Stereoscopy of the Coronal Magnetic Field}, \emph{Solar Phys.}, 290,
  2765--2789, \doi{10.1007/s11207-015-0791-z}

\bibitem[{\textbf{{Bai}}(1987)}]{bai_87}
{Bai}, T. (1987), {Distribution of flares on the sun - Superactive regions and
  active zones of 1980-1985}, \emph{Astrophys. J.}, 314, 795--807,
  \doi{10.1086/165105}

\bibitem[{\textbf{{Baliunas} et~al.}(1995)\textbf{{Baliunas}, {Donahue},
  {Soon}, {Horne}, {Frazer}, {Woodard-Eklund} et~al.}}]{baliunas_95}
{Baliunas}, S.~L., {Donahue}, R.~A., {Soon}, W.~H., {Horne}, J.~H., {Frazer},
  J., {Woodard-Eklund}, L., et~al. (1995), {Chromospheric variations in
  main-sequence stars}, \emph{Astrophys. J.}, 438, 269--287,
  \doi{10.1086/175072}

\bibitem[{\textbf{Balogh et~al.}(1992)\textbf{Balogh, Beek, Forsyth, Hedgecock,
  Marquedant, Smith et~al.}}]{balogh_92}
Balogh, A., Beek, T.~J., Forsyth, R.~J., Hedgecock, P.~C., Marquedant, R.~J.,
  Smith, E.~J., et~al. (1992), The magnetic field investigation on the ulysses
  mission instrumentation and preliminary scientific results, \emph{A\&AS}, 92,
  221

\bibitem[{\textbf{Bame et~al.}(1992)\textbf{Bame, McComas, Barraclough,
  Phillips, Sofaly, Chavez et~al.}}]{bame_92}
Bame, S., McComas, D., Barraclough, B., Phillips, J., Sofaly, K., Chavez, J.,
  et~al. (1992), The ulysses solar-wind plasma experiment, \emph{Astron.
  Astrophys. Supp. Series}, 92, 237

\bibitem[{\textbf{{Basu} and {Antia}}(2008)}]{basu_08}
{Basu}, S. and {Antia}, H.~M. (2008), {Helioseismology and solar abundances},
  \emph{Physics Reports}, 457, 217--283, \doi{10.1016/j.physrep.2007.12.002}

\bibitem[{\textbf{{Bird} et~al.}(1985)\textbf{{Bird}, {Volland}, {Howard},
  {Koomen}, {Michels}, {Sheeley} et~al.}}]{bird_85}
{Bird}, M.~K., {Volland}, H., {Howard}, R.~A., {Koomen}, M.~J., {Michels},
  D.~J., {Sheeley}, N.~R., Jr., et~al. (1985), {White-light and radio sounding
  observations of coronal transients}, \emph{Solar Phys.}, 98, 341--368,
  \doi{10.1007/BF00152465}

\bibitem[{\textbf{{Bogart} et~al.}(2015)\textbf{{Bogart}, {Baldner}, and
  {Basu}}}]{bogart_15}
{Bogart}, R.~S., {Baldner}, C.~S., and {Basu}, S. (2015), {Evolution of
  Near-surface Flows Inferred from High-resolution Ring-diagram Analysis},
  \emph{Astrophys. J.}, 807, 125, \doi{10.1088/0004-637X/807/2/125}

\bibitem[{\textbf{{Bosman} et~al.}(2012)\textbf{{Bosman}, {Bothmer},
  {Nistic{\`o}}, {Vourlidas}, {Howard}, and {Davies}}}]{bosman_12}
{Bosman}, E., {Bothmer}, V., {Nistic{\`o}}, G., {Vourlidas}, A., {Howard},
  R.~A., and {Davies}, J.~A. (2012), {Three-Dimensional Properties of Coronal
  Mass Ejections from STEREO/SECCHI Observations}, \emph{Solar Phys.}, 281,
  167--185, \doi{10.1007/s11207-012-0123-5}

\bibitem[{\textbf{{Casadei} et~al.}(2017)\textbf{{Casadei}, {Jeffrey}, and
  {Kontar}}}]{casadei_17}
{Casadei}, D., {Jeffrey}, N.~L.~S., and {Kontar}, E.~P. (2017), {Measuring
  X-ray anisotropy in solar flares. Prospective stereoscopic capabilities of
  STIX and MiSolFA}, \emph{aap}, 606, A2, \doi{10.1051/0004-6361/201730629}

\bibitem[{\textbf{{Castenmiller} et~al.}(1986)\textbf{{Castenmiller}, {Zwaan},
  and {van der Zalm}}}]{castenmiller_86}
{Castenmiller}, M.~J.~M., {Zwaan}, C., and {van der Zalm}, E.~B.~J. (1986),
  {Sunspot nests - Manifestations of sequences in magnetic activity},
  \emph{Solar Phys.}, 105, 237--255, \doi{10.1007/BF00172045}

\bibitem[{\textbf{{Chang} et~al.}(1997)\textbf{{Chang}, {Chou}, {Labonte}, and
  {TON Team}}}]{chang_97}
{Chang}, H.-K., {Chou}, D.-Y., {Labonte}, B., and {TON Team} (1997), {Ambient
  acoustic imaging in helioseismology}, \emph{Nature}, 389, 825--827,
  \doi{10.1038/39822}

\bibitem[{\textbf{{Charbonneau}}(2010)}]{charbonneau_10}
{Charbonneau}, P. (2010), {Dynamo Models of the Solar Cycle}, \emph{Living
  Reviews in Solar Physics}, 7, 3, \doi{10.12942/lrsp-2010-3}

\bibitem[{\textbf{{Christensen-Dalsgaard}}(2002)}]{christdals_02}
{Christensen-Dalsgaard}, J. (2002), {Helioseismology}, \emph{Reviews of Modern
  Physics}, 74, 1073--1129, \doi{10.1103/RevModPhys.74.1073}

\bibitem[{\textbf{{Christensen-Dalsgaard} and
  {Thompson}}(2007)}]{christdals_07}
{Christensen-Dalsgaard}, J. and {Thompson}, M.~J. (2007), {Observational
  results and issues concerning the tachocline}, in D.~W. {Hughes},
  R.~{Rosner}, and N.~O. {Weiss}, eds., The Solar Tachocline, 53

\bibitem[{\textbf{{Colaninno} et~al.}(2013)\textbf{{Colaninno}, {Vourlidas},
  and {Wu}}}]{colaninno_13}
{Colaninno}, R.~C., {Vourlidas}, A., and {Wu}, C.~C. (2013), {Quantitative
  comparison of methods for predicting the arrival of coronal mass ejections at
  Earth based on multiview imaging}, \emph{Journal of Geophysical Research
  (Space Physics)}, 118, 6866--6879, \doi{10.1002/2013JA019205}

\bibitem[{\textbf{{Colin}}(1980)}]{colin_80}
{Colin}, L. (1980), {The Pioneer Venus Program}, \emph{J. Geophys. Res.}, 85,
  7575--7598, \doi{10.1029/JA085iA13p07575}

\bibitem[{\textbf{{Davis} et~al.}(2009)\textbf{{Davis}, {Davies}, {Lockwood},
  {Rouillard}, {Eyles}, and {Harrison}}}]{davis_09}
{Davis}, C.~J., {Davies}, J.~A., {Lockwood}, M., {Rouillard}, A.~P., {Eyles},
  C.~J., and {Harrison}, R.~A. (2009), {Stereoscopic imaging of an
  Earth-impacting solar coronal mass ejection: A major milestone for the STEREO
  mission}, \emph{Geophys. Res. Lett.}, 36, L08102, \doi{10.1029/2009GL038021}

\bibitem[{\textbf{{De Rosa} et~al.}(2009)\textbf{{De Rosa}, {Schrijver},
  {Barnes}, {Leka}, {Lites}, {Aschwanden} et~al.}}]{Derosa09}
{De Rosa}, M.~L., {Schrijver}, C.~J., {Barnes}, G., {Leka}, K.~D., {Lites},
  B.~W., {Aschwanden}, M.~J., et~al. (2009), {A Critical Assessment of
  Nonlinear Force-Free Field Modeling of the Solar Corona for Active Region
  10953}, \emph{Astrophys. J.}, 696, 1780--1791,
  \doi{10.1088/0004-637X/696/2/1780}

\bibitem[{\textbf{{de Toma} et~al.}(2000)\textbf{{de Toma}, {White}, and
  {Harvey}}}]{detoma_00}
{de Toma}, G., {White}, O.~R., and {Harvey}, K.~L. (2000), {A Picture of Solar
  Minimum and the Onset of Solar Cycle 23. I. Global Magnetic Field Evolution},
  \emph{Astrophys. J.}, 529, 1101--1114, \doi{10.1086/308299}

\bibitem[{\textbf{DeForest et~al.}(2018)\textbf{DeForest, Howard, Velli, Viall,
  and Vourlidas}}]{deforest18}
DeForest, C., Howard, R.~A., Velli, M., Viall, N., and Vourlidas, A. (2018),
  The highly-structured outer solar corona, \emph{Astrophys. J.}

\bibitem[{\textbf{{Deforest} et~al.}(1997)\textbf{{Deforest}, {Hoeksema},
  {Gurman}, {Thompson}, {Plunkett}, {Howard} et~al.}}]{deforest_97}
{Deforest}, C.~E., {Hoeksema}, J.~T., {Gurman}, J.~B., {Thompson}, B.~J.,
  {Plunkett}, S.~P., {Howard}, R., et~al. (1997), {Polar Plume Anatomy: Results
  of a Coordinated Observation}, \emph{Solar Phys.}, 175, 393--410,
  \doi{10.1023/A:1004955223306}

\bibitem[{\textbf{{DeForest} et~al.}(2013)\textbf{{DeForest}, {Howard}, and
  {McComas}}}]{deforest_13}
{DeForest}, C.~E., {Howard}, T.~A., and {McComas}, D.~J. (2013), {Tracking
  Coronal Features from the Low Corona to Earth: A Quantitative Analysis of the
  2008 December 12 Coronal Mass Ejection}, \emph{Astrophys. J.}, 769, 43,
  \doi{10.1088/0004-637X/769/1/43}

\bibitem[{\textbf{{DeForest} et~al.}(2014)\textbf{{DeForest}, {Howard}, and
  {McComas}}}]{deforest_14}
{DeForest}, C.~E., {Howard}, T.~A., and {McComas}, D.~J. (2014), {Inbound Waves
  in the Solar Corona: A Direct Indicator of Alfv{\'e}n Surface Location},
  \emph{Astrophys. J.}, 787, 124, \doi{10.1088/0004-637X/787/2/124}

\bibitem[{\textbf{{DeForest} et~al.}(2016)\textbf{{DeForest}, {Howard}, {Webb},
  and {Davies}}}]{deforest_16}
{DeForest}, C.~E., {Howard}, T.~A., {Webb}, D.~F., and {Davies}, J.~A. (2016),
  {The utility of polarized heliospheric imaging for space weather monitoring},
  \emph{Space Weather}, 14, 32--49, \doi{10.1002/2015SW001286}

\bibitem[{\textbf{{Duvall} et~al.}(1993)\textbf{{Duvall}, {Jefferies},
  {Harvey}, and {Pomerantz}}}]{duvall_93}
{Duvall}, T.~L., Jr., {Jefferies}, S.~M., {Harvey}, J.~W., and {Pomerantz},
  M.~A. (1993), {Time-distance helioseismology}, \emph{Nature}, 362, 430--432,
  \doi{10.1038/362430a0}

\bibitem[{\textbf{{Egeland} et~al.}(2017)\textbf{{Egeland}, {Soon}, {Baliunas},
  {Hall}, {Pevtsov}, and {Bertello}}}]{egeland_17}
{Egeland}, R., {Soon}, W., {Baliunas}, S., {Hall}, J.~C., {Pevtsov}, A.~A., and
  {Bertello}, L. (2017), {VizieR Online Data Catalog: Calibrated solar S-index
  time series (Egeland+, 2017)}, \emph{VizieR Online Data Catalog}, 183

\bibitem[{\textbf{{Elsworth} et~al.}(1994)\textbf{{Elsworth}, {Howe}, {Isaak},
  {McLeod}, {Miller}, {New} et~al.}}]{elsworth_94}
{Elsworth}, Y., {Howe}, R., {Isaak}, G.~R., {McLeod}, C.~P., {Miller}, B.~A.,
  {New}, R., et~al. (1994), {Solar p-mode frequencies and their dependence on
  solar activity recent results from the BISON network}, \emph{Astrophys. J.},
  434, 801--806, \doi{10.1086/174783}

\bibitem[{\textbf{{Fan}}(2018)}]{fan_18}
{Fan}, Y. (2018), {MHD Simulation of a Prominence Eruption}, \emph{Astrophys.
  J.}

\bibitem[{\textbf{Forland et~al.}(2013)\textbf{Forland, Gibson, Dove,
  Rachmeler, and Fan}}]{forland_13}
Forland, B.~F., Gibson, S.~E., Dove, J.~B., Rachmeler, L.~A., and Fan, Y.
  (2013), Coronal cavity survey: morphological clues to eruptive magnetic
  topologies, \emph{Solar Phys., in press}

\bibitem[{\textbf{{Fox} et~al.}(2016)\textbf{{Fox}, {Velli}, {Bale}, {Decker},
  {Driesman}, {Howard} et~al.}}]{fox_16}
{Fox}, N.~J., {Velli}, M.~C., {Bale}, S.~D., {Decker}, R., {Driesman}, A.,
  {Howard}, R.~A., et~al. (2016), {The Solar Probe Plus Mission: Humanity's
  First Visit to Our Star}, \emph{Space Science Reviews}, 204, 7--48,
  \doi{10.1007/s11214-015-0211-6}

\bibitem[{\textbf{{Fulle} et~al.}(2007)\textbf{{Fulle}, {Leblanc}, {Harrison},
  {Davis}, {Eyles}, {Halain} et~al.}}]{fulle_07}
{Fulle}, M., {Leblanc}, F., {Harrison}, R.~A., {Davis}, C.~J., {Eyles}, C.~J.,
  {Halain}, J.~P., et~al. (2007), {Discovery of the Atomic Iron Tail of Comet
  MCNaught Using the Heliospheric Imager on STEREO}, \emph{Astrophys. J.
  Lett.}, 661, L93--L96, \doi{10.1086/518719}

\bibitem[{\textbf{{Gabriel} et~al.}(1997)\textbf{{Gabriel}, {Charra}, {Grec},
  {Robillot}, {Roca Cort{\'e}s}, {Turck-Chi{\`e}ze} et~al.}}]{gabriel_97}
{Gabriel}, A.~H., {Charra}, J., {Grec}, G., {Robillot}, J.-M., {Roca
  Cort{\'e}s}, T., {Turck-Chi{\`e}ze}, S., et~al. (1997), {Performance and
  Early Results from the GOLF Instrument Flown on the SOHO Mission},
  \emph{Solar Phys.}, 175, 207--226, \doi{10.1023/A:1004911408285}

\bibitem[{\textbf{{Gibson}}(2015)}]{gibson_15b}
{Gibson}, S. (2015), {Data-model comparison using FORWARD and CoMP}, in IAU
  Symposium, volume 305 of \emph{IAU Symposium}, volume 305 of \emph{IAU
  Symposium}, 245--250, \doi{10.1017/S1743921315004846}

\bibitem[{\textbf{Gibson et~al.}(2016)\textbf{Gibson, Kucera, White, Dove, Fan,
  Forland et~al.}}]{gibson_16}
Gibson, S., Kucera, T., White, S., Dove, J., Fan, Y., Forland, B., et~al.
  (2016), Forward: A toolset for multiwavelength coronal magnetometry,
  \emph{Front. Astron. Space Sci.}, 3

\bibitem[{\textbf{Gibson et~al.}(2009)\textbf{Gibson, Kozyra, de~Toma, Emery,
  Onsager, and Thompson}}]{gibhss}
Gibson, S.~E., Kozyra, J.~U., de~Toma, G., Emery, B.~A., Onsager, T., and
  Thompson, B.~J. (2009), If the sun is so quiet, why is the earth still
  ringing? a comparison of two solar minimum intervals, \emph{J. Geophys.
  Res.}, 114, A09105, \doi{10.1029/2009JA014342}

\bibitem[{\textbf{{Gizon} et~al.}(2010)\textbf{{Gizon}, {Birch}, and
  {Spruit}}}]{gizon_10}
{Gizon}, L., {Birch}, A.~C., and {Spruit}, H.~C. (2010), {Local
  Helioseismology: Three-Dimensional Imaging of the Solar Interior}, \emph{Ann.
  Rev. Astron. Astrophys.}, 48, 289--338,
  \doi{10.1146/annurev-astro-082708-101722}

\bibitem[{\textbf{Gloeckler et~al.}(1999)\textbf{Gloeckler, Geiss, Balsiger,
  Bedini, Cain, Fisher et~al.}}]{gloeck_92}
Gloeckler, G., Geiss, J., Balsiger, H., Bedini, P., Cain, J.~C., Fisher, J.,
  et~al. (1999), The solar wind ion composition spectrometer, \emph{Astron.
  Astrophys. Supp. Series}, 92, 267

\bibitem[{\textbf{{Goelzer} et~al.}(2014)\textbf{{Goelzer}, {Schwadron}, and
  {Smith}}}]{goelzer_14}
{Goelzer}, M.~L., {Schwadron}, N.~A., and {Smith}, C.~W. (2014), {An analysis
  of Alfv{\'e}n radius based on sunspot number from 1749 to today},
  \emph{Journal of Geophysical Research (Space Physics)}, 119, 115--120,
  \doi{10.1002/2013JA019420}

\bibitem[{\textbf{{Gopalswamy} et~al.}(2011)\textbf{{Gopalswamy}, {Davila},
  {St.~Cyr}, {Sittler}, {Auch{\`e}re}, {Duvall} et~al.}}]{gopal_11}
{Gopalswamy}, N., {Davila}, J.~M., {St.~Cyr}, O.~C., {Sittler}, E.~C.,
  {Auch{\`e}re}, F., {Duvall}, T.~L., et~al. (2011), {Earth-Affecting Solar
  Causes Observatory (EASCO): A potential International Living with a Star
  Mission from Sun-Earth L5}, \emph{Journal of Atmospheric and
  Solar-Terrestrial Physics}, 73, 658--663, \doi{10.1016/j.jastp.2011.01.013}

\bibitem[{\textbf{{Gurnett} et~al.}(1975)\textbf{{Gurnett}, {Anderson}, and
  {Odem}}}]{gurnett_75}
{Gurnett}, D.~A., {Anderson}, R.~R., and {Odem}, D.~L. (1975), {The University
  of Iowa Helios solar wind plasma wave experiment /E 5a/},
  \emph{Raumfahrtforschung}, 19, 245--247

\bibitem[{\textbf{{Hanasoge} et~al.}(2015)\textbf{{Hanasoge}, {Miesch}, {Roth},
  {Schou}, {Sch{\"u}ssler}, and {Thompson}}}]{hanasoge_15}
{Hanasoge}, S., {Miesch}, M.~S., {Roth}, M., {Schou}, J., {Sch{\"u}ssler}, M.,
  and {Thompson}, M.~J. (2015), {Solar Dynamics, Rotation, Convection and
  Overshoot}, \emph{Space Science Reviews}, 196, 79--99,
  \doi{10.1007/s11214-015-0144-0}

\bibitem[{\textbf{{Harvey} et~al.}(1996)\textbf{{Harvey}, {Hill}, {Hubbard},
  {Kennedy}, {Leibacher}, {Pintar} et~al.}}]{harvey_96}
{Harvey}, J.~W., {Hill}, F., {Hubbard}, R.~P., {Kennedy}, J.~R., {Leibacher},
  J.~W., {Pintar}, J.~A., et~al. (1996), {The Global Oscillation Network Group
  (GONG) Project}, \emph{Science}, 272, 1284--1286,
  \doi{10.1126/science.272.5266.1284}

\bibitem[{\textbf{Hassler et~al.}(2018{\natexlab{a}})\textbf{Hassler, Newmark,
  Velli, and Murphy}}]{hassler_18b}
Hassler, D., Newmark, J., Velli, M., and Murphy, N. (2018{\natexlab{a}}), in
  Proceedings of Solar Wind 15 Conference, Brussels, Belgium, 18-22 June 2018

\bibitem[{\textbf{Hassler et~al.}(2018{\natexlab{b}})\textbf{Hassler, Velli,
  Murphy, and Newmark}}]{hassler_18a}
Hassler, D., Velli, M., Murphy, N., and Newmark, J. (2018{\natexlab{b}}), in
  TESS Conference (21-24 May 2018)

\bibitem[{\textbf{{Hassler} et~al.}(1999)\textbf{{Hassler}, {Dammasch},
  {Lemaire}, {Brekke}, {Curdt}, {Mason} et~al.}}]{hassler_99}
{Hassler}, D.~M., {Dammasch}, I.~E., {Lemaire}, P., {Brekke}, P., {Curdt}, W.,
  {Mason}, H.~E., et~al. (1999), {Solar Wind Outflow and the Chromospheric
  Magnetic Network}, \emph{Science}, 283, 810,
  \doi{10.1126/science.283.5403.810}

\bibitem[{\textbf{{Hassler} et~al.}(1997)\textbf{{Hassler}, {Wilhelm},
  {Lemaire}, and {Sch{\"u}hle}}}]{hassler_97}
{Hassler}, D.~M., {Wilhelm}, K., {Lemaire}, P., and {Sch{\"u}hle}, U. (1997),
  {Observations of Polar Plumes with the SUMER Instrument on SOHO}, \emph{Solar
  Phys.}, 175, 375--391, \doi{10.1023/A:1004959324214}

\bibitem[{\textbf{{Hathaway}}(2015)}]{hathaway_15}
{Hathaway}, D.~H. (2015), {The Solar Cycle}, \emph{Living Reviews in Solar
  Physics}, 12, 4, \doi{10.1007/lrsp-2015-4}

\bibitem[{\textbf{{Hickmann} et~al.}(2015)\textbf{{Hickmann}, {Godinez},
  {Henney}, and {Arge}}}]{hickmann_15}
{Hickmann}, K.~S., {Godinez}, H.~C., {Henney}, C.~J., and {Arge}, C.~N. (2015),
  {Data Assimilation in the ADAPT Photospheric Flux Transport Model},
  \emph{Solar Phys.}, 290, 1105--1118, \doi{10.1007/s11207-015-0666-3}

\bibitem[{\textbf{{Hill}}(1988)}]{hill_88}
{Hill}, F. (1988), {Rings and trumpets - Three-dimensional power spectra of
  solar oscillations}, \emph{Astrophys. J.}, 333, 996--1013,
  \doi{10.1086/166807}

\bibitem[{\textbf{{Hoeksema} and {Scherrer}}(1986)}]{hoeksema_86}
{Hoeksema}, J.~T. and {Scherrer}, P.~H. (1986), {An atlas of photospheric
  magnetic field observations and computed coronal magnetic fields: 1976-1985},
  \emph{Solar Phys.}, 105, 205--211, \doi{10.1007/BF00156388}

\bibitem[{\textbf{Howard et~al.}(2008)\textbf{Howard, Moses, Vourlidas,
  Newmark, Socker, P. et~al.}}]{howard_08}
Howard, R.~A., Moses, J.~D., Vourlidas, A., Newmark, J.~S., Socker, D.~G., P.,
  P.~S., et~al. (2008), Sun earth connection coronal and heliospheric
  investigation (secchi), \emph{Space Science Reviews}, 136, 67

\bibitem[{\textbf{{Howe}}(2009)}]{howe_09}
{Howe}, R. (2009), {Solar Interior Rotation and its Variation}, \emph{Living
  Reviews in Solar Physics}, 6, 1, \doi{10.12942/lrsp-2009-1}

\bibitem[{\textbf{{Hurley} et~al.}(1992)\textbf{{Hurley}, {Sommer}, {Atteia},
  {Boer}, {Cline}, {Cotin} et~al.}}]{hurley_92}
{Hurley}, K., {Sommer}, M., {Atteia}, J.-L., {Boer}, M., {Cline}, T., {Cotin},
  F., et~al. (1992), {The solar X-ray/cosmic gamma-ray burst experiment aboard
  ULYSSES}, \emph{aaps}, 92, 401--410

\bibitem[{\textbf{{Isavnin} et~al.}(2014)\textbf{{Isavnin}, {Vourlidas}, and
  {Kilpua}}}]{isavnin_14}
{Isavnin}, A., {Vourlidas}, A., and {Kilpua}, E.~K.~J. (2014),
  {Three-Dimensional Evolution of Flux-Rope CMEs and Its Relation to the Local
  Orientation of the Heliospheric Current Sheet}, \emph{Solar Phys.}, 289,
  2141--2156, \doi{10.1007/s11207-013-0468-4}

\bibitem[{\textbf{Issautier et~al.}(2008)\textbf{Issautier, Le~Chat, N.,
  Moncuquet, Hoang, MacDowall et~al.}}]{issautier_08}
Issautier, K., Le~Chat, G., N., M.-V., Moncuquet, M., Hoang, S., MacDowall,
  R.~J., et~al. (2008), Electron properties of high-speed solar wind from polar
  coronal holes obtained by ulysses thermal noise spectroscopy: Not so dense,
  not so hot, \emph{Geophys. Res. Lett.}, L19101, \doi{0.1029/2008GL034912}

\bibitem[{\textbf{{Jackson}}(1985)}]{jackson_85}
{Jackson}, B.~V. (1985), {Imaging of coronal mass ejections by the HELIOS
  spacecraft}, \emph{Solar Phys.}, 100, 563--574, \doi{10.1007/BF00158445}

\bibitem[{\textbf{{Jensen} et~al.}(2013)\textbf{{Jensen}, {Bisi}, {Breen},
  {Heiles}, {Minter}, and {Vilas}}}]{jensen_13}
{Jensen}, E.~A., {Bisi}, M.~M., {Breen}, A.~R., {Heiles}, C., {Minter}, T., and
  {Vilas}, F. (2013), {Measurements of Faraday Rotation Through the Solar
  Corona During the 2009 Solar Minimum with the MESSENGER Spacecraft},
  \emph{Solar Phys.}, 285, 83--95, \doi{10.1007/s11207-012-0213-4}

\bibitem[{\textbf{{Kaiser} et~al.}(2008)\textbf{{Kaiser}, {Kucera}, {Davila},
  {St.~Cyr}, {Guhathakurta}, and {Christian}}}]{kaiser_08}
{Kaiser}, M.~L., {Kucera}, T.~A., {Davila}, J.~M., {St.~Cyr}, O.~C.,
  {Guhathakurta}, M., and {Christian}, E. (2008), {The STEREO Mission: An
  Introduction}, \emph{Space Science Reviews}, 136, 5--16,
  \doi{10.1007/s11214-007-9277-0}

\bibitem[{\textbf{{Kane} et~al.}(1998)\textbf{{Kane}, {Hurley}, {McTiernan},
  {Boer}, {Niel}, {Kosugi} et~al.}}]{kane_98}
{Kane}, S.~R., {Hurley}, K., {McTiernan}, J.~M., {Boer}, M., {Niel}, M.,
  {Kosugi}, T., et~al. (1998), {Stereoscopic Observations of Solar Hard X-Ray
  Flares Made by ULYSSES and YOHKOH}, \emph{apj}, 500, 1003--1008,
  \doi{10.1086/305738}

\bibitem[{\textbf{{Katsikas} et~al.}(2010)\textbf{{Katsikas}, {Exarhos}, and
  {Moussas}}}]{katsikas_10}
{Katsikas}, V., {Exarhos}, G., and {Moussas}, X. (2010), {Study of the solar
  Slow Sonic, Alfv{\'e}n and Fast Magnetosonic transition surfaces},
  \emph{Advances in Space Research}, 46, 382--390,
  \doi{10.1016/j.asr.2010.05.003}

\bibitem[{\textbf{{Kosovichev}}(1996)}]{kosovichev_96}
{Kosovichev}, A.~G. (1996), {Tomographic Imaging of the Sun's Interior},
  \emph{Astrophys. J. Lett.}, 461, L55, \doi{10.1086/309989}

\bibitem[{\textbf{{Kosovichev} et~al.}(1997)\textbf{{Kosovichev}, {Schou},
  {Scherrer}, {Bogart}, {Bush}, {Hoeksema} et~al.}}]{kosovichev_97}
{Kosovichev}, A.~G., {Schou}, J., {Scherrer}, P.~H., {Bogart}, R.~S., {Bush},
  R.~I., {Hoeksema}, J.~T., et~al. (1997), {Structure and Rotation of the Solar
  Interior: Initial Results from the MDI Medium-L Program}, \emph{Solar Phys.},
  170, 43--61, \doi{10.1023/A:1004949311268}

\bibitem[{\textbf{{Kosugi} et~al.}(1991)\textbf{{Kosugi}, {Makishima},
  {Murakami}, {Sakao}, {Dotani}, {Inda} et~al.}}]{kosugi_91}
{Kosugi}, T., {Makishima}, K., {Murakami}, T., {Sakao}, T., {Dotani}, T.,
  {Inda}, M., et~al. (1991), {The Hard X-ray Telescope (HXT) for the SOLAR-A
  Mission}, \emph{solphys}, 136, 17--36, \doi{10.1007/BF00151693}

\bibitem[{\textbf{{Kramar} et~al.}(2016)\textbf{{Kramar}, {Airapetian}, and
  {Lin}}}]{kramar_16b}
{Kramar}, M., {Airapetian}, V., and {Lin}, H. (2016), {3D Global Coronal
  Density Structure and Associated Magnetic Field near Solar Maximum},
  \emph{Frontiers in Astronomy and Space Sciences}, 3, 25,
  \doi{10.3389/fspas.2016.00025}

\bibitem[{\textbf{{Kramar} et~al.}(2006)\textbf{{Kramar}, {Inhester}, and
  {Solanki}}}]{kramar_06}
{Kramar}, M., {Inhester}, B., and {Solanki}, S.~K. (2006), {Vector tomography
  for the coronal magnetic field. I. Longitudinal Zeeman effect measurements},
  \emph{aap}, 456, 665--673, \doi{10.1051/0004-6361:20064865}

\bibitem[{\textbf{{Lavraud} et~al.}(2016)\textbf{{Lavraud}, {Liu}, {Segura},
  {He}, {Qin}, {Temmer} et~al.}}]{lavraud_16}
{Lavraud}, B., {Liu}, Y., {Segura}, K., {He}, J., {Qin}, G., {Temmer}, M.,
  et~al. (2016), {A small mission concept to the Sun-Earth Lagrangian L5 point
  for innovative solar, heliospheric and space weather science}, \emph{Journal
  of Atmospheric and Solar-Terrestrial Physics}, 146, 171--185,
  \doi{10.1016/j.jastp.2016.06.004}

\bibitem[{\textbf{{Leinert} et~al.}(1981)\textbf{{Leinert}, {Richter}, {Pitz},
  and {Planck}}}]{leinert_81}
{Leinert}, C., {Richter}, I., {Pitz}, E., and {Planck}, B. (1981), {The
  zodiacal light from 1.0 to 0.3 A.U. as observed by the HELIOS space probes},
  \emph{Astron. Astrophys.}, 103, 177--188

\bibitem[{\textbf{Liewer et~al.}(2009)\textbf{Liewer, Alexander, Ayon,
  Kosovichev, Mewaldt, Socker et~al.}}]{liewer_09}
Liewer, P.~C., Alexander, D., Ayon, J., Kosovichev, A., Mewaldt, R., Socker,
  D., et~al. (2009), Solar polar imager: Observing solar activity from a new
  perspective, in M.~S. {Allen}, ed., NASA Space Science Vision Missions,
  volume 224 (American Institute of Aeronautics and Astronautics, Progress in
  Astronautics and Aeronautics), volume 224, 1

\bibitem[{\textbf{{Liewer} et~al.}(2017)\textbf{{Liewer}, {Qiu}, and
  {Lindsey}}}]{liewer_17}
{Liewer}, P.~C., {Qiu}, J., and {Lindsey}, C. (2017), {Comparison of
  Helioseismic Far-Side Active Region Detections with STEREO Far-Side EUV
  Observations of Solar Activity}, \emph{Solar Phys.}, 292, 146,
  \doi{10.1007/s11207-017-1159-3}

\bibitem[{\textbf{{Lindsey} and {Braun}}(1997)}]{lindsey_97}
{Lindsey}, C. and {Braun}, D.~C. (1997), {Helioseismic Holography},
  \emph{Astrophys. J.}, 485, 895--903, \doi{10.1086/304445}

\bibitem[{\textbf{{Linker} et~al.}(2017)\textbf{{Linker}, {Caplan}, {Downs},
  {Riley}, {Mikic}, {Lionello} et~al.}}]{linker_17}
{Linker}, J.~A., {Caplan}, R.~M., {Downs}, C., {Riley}, P., {Mikic}, Z.,
  {Lionello}, R., et~al. (2017), {The Open Flux Problem}, \emph{Astrophys. J.},
  848, 70, \doi{10.3847/1538-4357/aa8a70}

\bibitem[{\textbf{{L{\"o}ptien} et~al.}(2015)\textbf{{L{\"o}ptien}, {Birch},
  {Gizon}, {Schou}, {Appourchaux}, {Blanco Rodr{\'{\i}}guez}
  et~al.}}]{loeptien_15}
{L{\"o}ptien}, B., {Birch}, A.~C., {Gizon}, L., {Schou}, J., {Appourchaux}, T.,
  {Blanco Rodr{\'{\i}}guez}, J., et~al. (2015), {Helioseismology with Solar
  Orbiter}, \emph{Space Science Reviews}, 196, 251--283,
  \doi{10.1007/s11214-014-0065-3}

\bibitem[{\textbf{{Lugaz} et~al.}(2017)\textbf{{Lugaz}, {Temmer}, {Wang}, and
  {Farrugia}}}]{lugaz_17}
{Lugaz}, N., {Temmer}, M., {Wang}, Y., and {Farrugia}, C.~J. (2017), {The
  Interaction of Successive Coronal Mass Ejections: A Review}, \emph{Solar
  Phys.}, 292, 64, \doi{10.1007/s11207-017-1091-6}

\bibitem[{\textbf{{Mackay} and {Yeates}}(2012)}]{mackay_12}
{Mackay}, D.~H. and {Yeates}, A.~R. (2012), {The Sun's Global Photospheric and
  Coronal Magnetic Fields: Observations and Models}, \emph{Living Reviews in
  Solar Physics}, 9, 6, \doi{10.12942/lrsp-2012-6}

\bibitem[{\textbf{{Mackay} et~al.}(2016)\textbf{{Mackay}, {Yeates}, and
  {Bocquet}}}]{mackay_16}
{Mackay}, D.~H., {Yeates}, A.~R., and {Bocquet}, F.-X. (2016), {Impact of an L5
  Magnetograph on Nonpotential Solar Global Magnetic Field Modeling},
  \emph{Astrophys. J.}, 825, 131, \doi{10.3847/0004-637X/825/2/131}

\bibitem[{\textbf{{Malanushenko} et~al.}(2014)\textbf{{Malanushenko},
  {Schrijver}, {DeRosa}, and {Wheatland}}}]{malanushenko_14}
{Malanushenko}, A., {Schrijver}, C.~J., {DeRosa}, M.~L., and {Wheatland}, M.~S.
  (2014), {Using Coronal Loops to Reconstruct the Magnetic Field of an Active
  Region before and after a Major Flare}, \emph{APJ}, 783, 102,
  \doi{10.1088/0004-637X/783/2/102}

\bibitem[{\textbf{{McComas} et~al.}(1998)\textbf{{McComas}, {Bame},
  {Barraclough}, {Feldman}, {Funsten}, {Gosling} et~al.}}]{mccomas_98}
{McComas}, D.~J., {Bame}, S.~J., {Barraclough}, B.~L., {Feldman}, W.~C.,
  {Funsten}, H.~O., {Gosling}, J.~T., et~al. (1998), {Ulysses' return to the
  slow solar wind}, \emph{Geophys. Res. Lett.}, 25, 1--4,
  \doi{10.1029/97GL03444}

\bibitem[{\textbf{{McComas} et~al.}(2003)\textbf{{McComas}, {Elliott},
  {Schwadron}, {Gosling}, {Skoug}, and {Goldstein}}}]{mccomas_03}
{McComas}, D.~J., {Elliott}, H.~A., {Schwadron}, N.~A., {Gosling}, J.~T.,
  {Skoug}, R.~M., and {Goldstein}, B.~E. (2003), {The three-dimensional solar
  wind around solar maximum}, \emph{Geophys. Res. Lett.}, 30, 1517,
  \doi{10.1029/2003GL017136}

\bibitem[{\textbf{{McDonald} et~al.}(1977)\textbf{{McDonald}, {Trainor}, {Lal},
  {Van Hollebeke}, and {Webber}}}]{mcdonald_77}
{McDonald}, F.~B., {Trainor}, J.~H., {Lal}, N., {Van Hollebeke}, M.~A.~I., and
  {Webber}, W.~R. (1977), {Observations of galactic cosmic-ray energy spectra
  between 1 and 9 AU}, \emph{Astrophys. J.}, 216, 930--939,
  \doi{10.1086/155537}

\bibitem[{\textbf{McIntosh}(2003)}]{patmap}
McIntosh, P.~S. (2003), Patterns and dynamics of solar magnetic fields and hei
  coronal holes in cycle 23, 807

\bibitem[{\textbf{{McIntosh} et~al.}(1976)\textbf{{McIntosh}, {Krieger},
  {Nolte}, and {Vaiana}}}]{mcintosh_76}
{McIntosh}, P.~S., {Krieger}, A.~S., {Nolte}, J.~T., and {Vaiana}, G. (1976),
  {Association of X-ray arches with chromospheric neutral lines}, \emph{Solar
  Phys.}, 49, 57--77, \doi{10.1007/BF00221485}

\bibitem[{\textbf{{McIntosh} et~al.}(2006)\textbf{{McIntosh}, {Davey}, and
  {Hassler}}}]{mcintosh_06}
{McIntosh}, S.~W., {Davey}, A.~R., and {Hassler}, D.~M. (2006), {Simple
  Magnetic Flux Balance as an Indicator of Ne VIII Doppler Velocity
  Partitioning in an Equatorial Coronal Hole}, \emph{Astrophys. J. Lett.}, 644,
  L87--L91, \doi{10.1086/505488}

\bibitem[{\textbf{{McIntosh} et~al.}(2015)\textbf{{McIntosh}, {Leamon},
  {Krista}, {Title}, {Hudson}, {Riley} et~al.}}]{mcintosh_15}
{McIntosh}, S.~W., {Leamon}, R.~J., {Krista}, L.~D., {Title}, A.~M., {Hudson},
  H.~S., {Riley}, P., et~al. (2015), {The solar magnetic activity band
  interaction and instabilities that shape quasi-periodic variability},
  \emph{Nature Communications}, 6, 6491, \doi{10.1038/ncomms7491}

\bibitem[{\textbf{{Miesch}}(2005)}]{miesch_05}
{Miesch}, M.~S. (2005), {Large-Scale Dynamics of the Convection Zone and
  Tachocline}, \emph{Living Reviews in Solar Physics}, 2, 1,
  \doi{10.12942/lrsp-2005-1}

\bibitem[{\textbf{{M{\"u}ller} et~al.}(2013)\textbf{{M{\"u}ller}, {Marsden},
  {St.~Cyr}, and {Gilbert}}}]{mueller_13}
{M{\"u}ller}, D., {Marsden}, R.~G., {St.~Cyr}, O.~C., and {Gilbert}, H.~R.
  (2013), {Solar Orbiter . Exploring the Sun-Heliosphere Connection},
  \emph{Solar Phys.}, 285, 25--70, \doi{10.1007/s11207-012-0085-7}

\bibitem[{\textbf{{Neugebauer} et~al.}(2007)\textbf{{Neugebauer}, {Gloeckler},
  {Gosling}, {Rees}, {Skoug}, {Goldstein} et~al.}}]{neugebauer_07}
{Neugebauer}, M., {Gloeckler}, G., {Gosling}, J.~T., {Rees}, A., {Skoug}, R.,
  {Goldstein}, B.~E., et~al. (2007), {Encounter of the Ulysses Spacecraft with
  the Ion Tail of Comet MCNaught}, \emph{Astrophys. J.}, 667, 1262--1266,
  \doi{10.1086/521019}

\bibitem[{\textbf{Newmark}(2018)}]{newmark_18}
Newmark, J. (2018), in TESS Conference (21-24 May 2018)

\bibitem[{\textbf{{Olmedo} et~al.}(2012)\textbf{{Olmedo}, {Vourlidas}, {Zhang},
  and {Cheng}}}]{olmedo_12}
{Olmedo}, O., {Vourlidas}, A., {Zhang}, J., and {Cheng}, X. (2012), {Secondary
  Waves and/or the ``Reflection'' from and ``Transmission'' through a Coronal
  Hole of an Extreme Ultraviolet Wave Associated with the 2011 February 15 X2.2
  Flare Observed with SDO/AIA and STEREO/EUVI}, \emph{Astrophys. J.}, 756, 143,
  \doi{10.1088/0004-637X/756/2/143}

\bibitem[{\textbf{{Patsourakos} et~al.}(2016)\textbf{{Patsourakos},
  {Georgoulis}, {Vourlidas}, {Nindos}, {Sarris}, {Anagnostopoulos}
  et~al.}}]{patsour_16}
{Patsourakos}, S., {Georgoulis}, M.~K., {Vourlidas}, A., {Nindos}, A.,
  {Sarris}, T., {Anagnostopoulos}, G., et~al. (2016), {The Major Geoeffective
  Solar Eruptions of 2012 March 7: Comprehensive Sun-to-Earth Analysis},
  \emph{apj}, 817, 14, \doi{10.3847/0004-637X/817/1/14}

\bibitem[{\textbf{Petrie and Low}(2005)}]{petrie}
Petrie, G. J.~D. and Low, B.~C. (2005), The dynamical consequences of
  spontaneous current sheets in quiescent prominences, \emph{Astrophys. J.
  Supp.}, 159, 288

\bibitem[{\textbf{{Pevtsov} et~al.}(2016)\textbf{{Pevtsov}, {Bertello},
  {MacNeice}, and {Petrie}}}]{pevtsov_16}
{Pevtsov}, A.~A., {Bertello}, L., {MacNeice}, P., and {Petrie}, G. (2016),
  {What if we had a magnetograph at Lagrangian L5?}, \emph{Space Weather}, 14,
  1026--1031, \doi{10.1002/2016SW001471}

\bibitem[{\textbf{{Rieutord} and {Rincon}}(2010)}]{rieutord_10}
{Rieutord}, M. and {Rincon}, F. (2010), {The Sun's Supergranulation},
  \emph{Living Reviews in Solar Physics}, 7, 2, \doi{10.12942/lrsp-2010-2}

\bibitem[{\textbf{{Riley} et~al.}(2014)\textbf{{Riley}, {Ben-Nun}, {Linker},
  {Mikic}, {Svalgaard}, {Harvey} et~al.}}]{riley_14}
{Riley}, P., {Ben-Nun}, M., {Linker}, J.~A., {Mikic}, Z., {Svalgaard}, L.,
  {Harvey}, J., et~al. (2014), {A Multi-Observatory Inter-Comparison of
  Line-of-Sight Synoptic Solar Magnetograms}, \emph{solphys}, 289, 769--792,
  \doi{10.1007/s11207-013-0353-1}

\bibitem[{\textbf{{Riley} et~al.}(2012)\textbf{{Riley}, {Linker}, {Lionello},
  and {Mikic}}}]{rileyetal_12}
{Riley}, P., {Linker}, J.~A., {Lionello}, R., and {Mikic}, Z. (2012),
  {Corotating interaction regions during the recent solar minimum: The power
  and limitations of global MHD modeling}, \emph{Journal of Atmospheric and
  Solar-Terrestrial Physics}, 83, 1--10, \doi{10.1016/j.jastp.2011.12.013}

\bibitem[{\textbf{{Robbrecht} et~al.}(2009)\textbf{{Robbrecht}, {Patsourakos},
  and {Vourlidas}}}]{robbrecht_09}
{Robbrecht}, E., {Patsourakos}, S., and {Vourlidas}, A. (2009), {No Trace Left
  Behind: STEREO Observation of a Coronal Mass Ejection Without Low Coronal
  Signatures}, \emph{Astrophys. J.}, 701, 283--291,
  \doi{10.1088/0004-637X/701/1/283}

\bibitem[{\textbf{{Rosenbauer} et~al.}(1977)\textbf{{Rosenbauer}, {Schwenn},
  {Marsch}, {Meyer}, {Miggenrieder}, {Montgomery} et~al.}}]{rosenbauer_77}
{Rosenbauer}, H., {Schwenn}, R., {Marsch}, E., {Meyer}, B., {Miggenrieder}, H.,
  {Montgomery}, M.~D., et~al. (1977), {A survey on initial results of the
  HELIOS plasma experiment}, \emph{Journal of Geophysics Zeitschrift
  Geophysik}, 42, 561--580

\bibitem[{\textbf{{Rouillard} et~al.}(2009)\textbf{{Rouillard}, {Savani},
  {Davies}, {Lavraud}, {Forsyth}, {Morley} et~al.}}]{rouillard_09}
{Rouillard}, A.~P., {Savani}, N.~P., {Davies}, J.~A., {Lavraud}, B., {Forsyth},
  R.~J., {Morley}, S.~K., et~al. (2009), {A Multispacecraft Analysis of a
  Small-Scale Transient Entrained by Solar Wind Streams}, \emph{Solar Phys.},
  256, 307--326, \doi{10.1007/s11207-009-9329-6}

\bibitem[{\textbf{{Sachdeva} et~al.}(2017)\textbf{{Sachdeva}, {Subramanian},
  {Vourlidas}, and {Bothmer}}}]{sachdeva_17}
{Sachdeva}, N., {Subramanian}, P., {Vourlidas}, A., and {Bothmer}, V. (2017),
  {CME Dynamics Using STEREO and LASCO Observations: The Relative Importance of
  Lorentz Forces and Solar Wind Drag}, \emph{Solar Phys.}, 292, 118,
  \doi{10.1007/s11207-017-1137-9}

\bibitem[{\textbf{{Savani} et~al.}(2013)\textbf{{Savani}, {Vourlidas},
  {Pulkkinen}, {Nieves-Chinchilla}, {Lavraud}, and {Owens}}}]{savani_13}
{Savani}, N.~P., {Vourlidas}, A., {Pulkkinen}, A., {Nieves-Chinchilla}, T.,
  {Lavraud}, B., and {Owens}, M.~J. (2013), {Tracking the momentum flux of a
  CME and quantifying its influence on geomagnetically induced currents at
  Earth}, \emph{Space Weather}, 11, 245--261, \doi{10.1002/swe.20038}

\bibitem[{\textbf{{Savani} et~al.}(2017)\textbf{{Savani}, {Vourlidas},
  {Richardson}, {Szabo}, {Thompson}, {Pulkkinen} et~al.}}]{savani_17}
{Savani}, N.~P., {Vourlidas}, A., {Richardson}, I.~G., {Szabo}, A., {Thompson},
  B.~J., {Pulkkinen}, A., et~al. (2017), {Predicting the magnetic vectors
  within coronal mass ejections arriving at Earth: 2. Geomagnetic response},
  \emph{Space Weather}, 15, 441--461, \doi{10.1002/2016SW001458}

\bibitem[{\textbf{{Savani} et~al.}(2015)\textbf{{Savani}, {Vourlidas}, {Szabo},
  {Mays}, {Richardson}, {Thompson} et~al.}}]{savani_15}
{Savani}, N.~P., {Vourlidas}, A., {Szabo}, A., {Mays}, M.~L., {Richardson},
  I.~G., {Thompson}, B.~J., et~al. (2015), {Predicting the magnetic vectors
  within coronal mass ejections arriving at Earth: 1. Initial architecture},
  \emph{Space Weather}, 13, 374--385, \doi{10.1002/2015SW001171}

\bibitem[{\textbf{Savcheva et~al.}(2013)\textbf{Savcheva, McKillop, Hanson,
  McCauly, and DeLuca}}]{Savcheva13b}
Savcheva, A., McKillop, S., Hanson, E., McCauly, P., and DeLuca, E. (2013),
  Observed properties of sigmoidal regions and sigmoid evolutionary histories,
  \emph{Solar Physics}, in preparation

\bibitem[{\textbf{{Schou} and {Bogart}}(1998)}]{schou_98}
{Schou}, J. and {Bogart}, R.~S. (1998), {Flows and Horizontal Displacements
  from Ring Diagrams}, \emph{Astrophys. J. Lett.}, 504, L131--L134,
  \doi{10.1086/311575}

\bibitem[{\textbf{Schrijver and DeRosa}(2003)}]{schrijver_03}
Schrijver, C. and DeRosa, M. (2003), \emph{Solar Phys.}, 212, 165

\bibitem[{\textbf{{Schrijver} and {Title}}(2011)}]{schrijver_11}
{Schrijver}, C.~J. and {Title}, A.~M. (2011), {Long-range magnetic couplings
  between solar flares and coronal mass ejections observed by SDO and STEREO},
  \emph{Journal of Geophysical Research (Space Physics)}, 116, 15, A04108,
  \doi{10.1029/2010JA016224}

\bibitem[{\textbf{{Sekii} et~al.}(2015)\textbf{{Sekii}, {Appourchaux}, {Fleck},
  and {Turck-Chi{\`e}ze}}}]{sekii_15}
{Sekii}, T., {Appourchaux}, T., {Fleck}, B., and {Turck-Chi{\`e}ze}, S. (2015),
  {Future Mission Concepts for Helioseismology}, \emph{Space Science Reviews},
  196, 285--302, \doi{10.1007/s11214-015-0142-2}

\bibitem[{\textbf{{Semel} and {Skumanich}}(1998)}]{semel_98}
{Semel}, M. and {Skumanich}, A. (1998), {An ambiguity-free determination of
  J\_Z in solar active regions}, \emph{Astron. Astrophys.}, 331, 383--391

\bibitem[{\textbf{{Shapiro} et~al.}(2014)\textbf{{Shapiro}, {Solanki},
  {Krivova}, {Schmutz}, {Ball}, {Knaack} et~al.}}]{shapiro_14}
{Shapiro}, A.~I., {Solanki}, S.~K., {Krivova}, N.~A., {Schmutz}, W.~K., {Ball},
  W.~T., {Knaack}, R., et~al. (2014), {Variability of Sun-like stars:
  reproducing observed photometric trends}, \emph{Astron. Astrophys.}, 569,
  A38, \doi{10.1051/0004-6361/201323086}

\bibitem[{\textbf{{Shapiro} et~al.}(2016)\textbf{{Shapiro}, {Solanki},
  {Krivova}, {Yeo}, and {Schmutz}}}]{shapiro_16}
{Shapiro}, A.~I., {Solanki}, S.~K., {Krivova}, N.~A., {Yeo}, K.~L., and
  {Schmutz}, W.~K. (2016), {Are solar brightness variations faculae- or
  spot-dominated?}, \emph{Astron. Astrophys.}, 589, A46,
  \doi{10.1051/0004-6361/201527527}

\bibitem[{\textbf{{Sheeley} et~al.}(2008)\textbf{{Sheeley}, {Herbst},
  {Palatchi}, {Wang}, {Howard}, {Moses} et~al.}}]{sheeleyetal_08}
{Sheeley}, N.~R., Jr., {Herbst}, A.~D., {Palatchi}, C.~A., {Wang}, Y.-M.,
  {Howard}, R.~A., {Moses}, J.~D., et~al. (2008), {SECCHI Observations of the
  Sun's Garden-Hose Density Spiral}, \emph{Astrophys. J. Lett.}, 674, L109,
  \doi{10.1086/529020}

\bibitem[{\textbf{{Solomon} et~al.}(2001)\textbf{{Solomon}, {McNutt}, {Gold},
  {Santo}, and {MESSENGER Team}}}]{solomon_01}
{Solomon}, S.~C., {McNutt}, R.~L., Jr., {Gold}, R.~E., {Santo}, A.~G., and
  {MESSENGER Team} (2001), {The MESSENGER Mission to Mercury}, in
  M.~{Robinbson} and G.~J. {Taylor}, eds., Workshop on Mercury: Space
  Environment, Surface, and Interior, volume 1097, volume 1097, 94

\bibitem[{\textbf{{Tasnim} and {Cairns}}(2016)}]{tasnim_16}
{Tasnim}, S. and {Cairns}, I.~H. (2016), {An equatorial solar wind model with
  angular momentum conservation and nonradial magnetic fields and flow
  velocities at an inner boundary}, \emph{Journal of Geophysical Research
  (Space Physics)}, 121, 4966--4984, \doi{10.1002/2016JA022725}

\bibitem[{\textbf{{Thernisien} et~al.}(2009)\textbf{{Thernisien}, {Vourlidas},
  and {Howard}}}]{thernisien_09}
{Thernisien}, A., {Vourlidas}, A., and {Howard}, R.~A. (2009), {Forward
  Modeling of Coronal Mass Ejections Using STEREO/SECCHI Data}, \emph{Solar
  Phys.}, 256, 111--130, \doi{10.1007/s11207-009-9346-5}

\bibitem[{\textbf{{Thompson} and {Myers}}(2009)}]{thompson_09}
{Thompson}, B.~J. and {Myers}, D.~C. (2009), {A Catalog of Coronal ``EIT Wave''
  Transients}, \emph{Astrophys. J. Supp.}, 183, 225--243,
  \doi{10.1088/0067-0049/183/2/225}

\bibitem[{\textbf{{Thompson} et~al.}(2003)\textbf{{Thompson},
  {Christensen-Dalsgaard}, {Miesch}, and {Toomre}}}]{thompson_03}
{Thompson}, M.~J., {Christensen-Dalsgaard}, J., {Miesch}, M.~S., and {Toomre},
  J. (2003), {The Internal Rotation of the Sun}, \emph{Ann. Rev. Astron.
  Astrophys.}, 41, 599--643, \doi{10.1146/annurev.astro.41.011802.094848}

\bibitem[{\textbf{{Thompson}}(2006)}]{thompson_06}
{Thompson}, W.~T. (2006), {Coordinate systems for solar image data},
  \emph{Astron. Astrophys.}, 449, 791--803, \doi{10.1051/0004-6361:20054262}

\bibitem[{\textbf{{Titov} et~al.}(2012)\textbf{{Titov}, {Mikic},
  {T{\"o}r{\"o}k}, {Linker}, and {Panasenco}}}]{Titov12}
{Titov}, V.~S., {Mikic}, Z., {T{\"o}r{\"o}k}, T., {Linker}, J.~A., and
  {Panasenco}, O. (2012), {2010 August 1-2 Sympathetic Eruptions. I. Magnetic
  Topology of the Source-surface Background Field}, \emph{Astrophys. J.}, 759,
  70, \doi{10.1088/0004-637X/759/1/70}

\bibitem[{\textbf{{Titov} et~al.}(2017)\textbf{{Titov}, {Miki{\'c}},
  {T{\"o}r{\"o}k}, {Linker}, and {Panasenco}}}]{titov_17}
{Titov}, V.~S., {Miki{\'c}}, Z., {T{\"o}r{\"o}k}, T., {Linker}, J.~A., and
  {Panasenco}, O. (2017), {2010 August 1-2 Sympathetic Eruptions. II. Magnetic
  Topology of the MHD Background Field}, \emph{Astrophys. J.}, 845, 141,
  \doi{10.3847/1538-4357/aa81ce}

\bibitem[{\textbf{{T{\'o}th} et~al.}(2005)\textbf{{T{\'o}th}, {Sokolov},
  {Gombosi}, {Chesney}, {Clauer}, {de Zeeuw} et~al.}}]{toth_05}
{T{\'o}th}, G., {Sokolov}, I.~V., {Gombosi}, T.~I., {Chesney}, D.~R., {Clauer},
  C.~R., {de Zeeuw}, D.~L., et~al. (2005), {Space Weather Modeling Framework: A
  new tool for the space science community}, \emph{Journal of Geophysical
  Research (Space Physics)}, 110, 9, A12226, \doi{10.1029/2005JA011126}

\bibitem[{\textbf{{Trichas} et~al.}(2015)\textbf{{Trichas}, {Gibbs},
  {Harrison}, {Green}, {Eastwood}, {Bentley} et~al.}}]{trichas_15}
{Trichas}, M., {Gibbs}, M., {Harrison}, R., {Green}, L., {Eastwood}, J.,
  {Bentley}, B., et~al. (2015), {Carrington-L5: The UK/US Operational Space
  Weather Monitoring Mission}, \emph{Hipparchos, vol.~2, Issue 12, pp.~25 -
  31}, 2, 12, 25--31

\bibitem[{\textbf{{Tsuneta} et~al.}(2008)\textbf{{Tsuneta}, {Ichimoto},
  {Katsukawa}, {Lites}, {Matsuzaki}, {Nagata} et~al.}}]{tsuneta_08}
{Tsuneta}, S., {Ichimoto}, K., {Katsukawa}, Y., {Lites}, B.~W., {Matsuzaki},
  K., {Nagata}, S., et~al. (2008), {The Magnetic Landscape of the Sun's Polar
  Region}, \emph{Astrophys. J.}, 688, 1374-1381, \doi{10.1086/592226}

\bibitem[{\textbf{{Ugarte-Urra} et~al.}(2015)\textbf{{Ugarte-Urra}, {Upton},
  {Warren}, and {Hathaway}}}]{ugarte_15}
{Ugarte-Urra}, I., {Upton}, L., {Warren}, H.~P., and {Hathaway}, D.~H. (2015),
  {Magnetic Flux Transport and the Long-term Evolution of Solar Active
  Regions}, \emph{Astrophys. J.}, 815, 90, \doi{10.1088/0004-637X/815/2/90}

\bibitem[{\textbf{Vasquez et~al.}(2011)\textbf{Vasquez, Huang, Manchester, and
  Frazin}}]{vasquez_11}
Vasquez, A., Huang, Z., Manchester, W.~B., and Frazin, R.~A. (2011), The whi
  corona from differential emission measure tomography, \emph{Solar Phys.},
  \doi{10.1007/s11207-010-9706-1}

\bibitem[{\textbf{{Verdini} et~al.}(2009)\textbf{{Verdini}, {Velli}, and
  {Buchlin}}}]{verdini_09}
{Verdini}, A., {Velli}, M., and {Buchlin}, E. (2009), {Turbulence in the
  Sub-Alfv{\'e}nic Solar Wind Driven by Reflection of Low-Frequency Alfv{\'e}n
  Waves}, \emph{apjl}, 700, L39--L42, \doi{10.1088/0004-637X/700/1/L39}

\bibitem[{\textbf{{Vourlidas}}(2015)}]{vourlidas_15}
{Vourlidas}, A. (2015), {Mission to the Sun-Earth L$_{5}$ Lagrangian Point: An
  Optimal Platform for Space Weather Research}, \emph{Space Weather}, 13,
  197--201, \doi{10.1002/2015SW001173}

\bibitem[{\textbf{{Vourlidas}}(2017)}]{vourlidas_18}
{Vourlidas}, A. (2017), {Solar Polar Diamond Explorer (SPDEx): Understanding
  the Origins of Solar Activity Using a New Perspective}, \emph{White paper
  submitted in response to ideas for the Next Generation Solar Physics Mission
  Concepts}

\bibitem[{\textbf{{Wang} et~al.}(2004)\textbf{{Wang}, {Shen}, {Wang}, and
  {Ye}}}]{wang_04}
{Wang}, Y., {Shen}, C., {Wang}, S., and {Ye}, P. (2004), {Deflection of coronal
  mass ejection in the interplanetary medium}, \emph{Solar Phys.}, 222,
  329--343, \doi{10.1023/B:SOLA.0000043576.21942.aa}

\bibitem[{\textbf{Wang and Sheeley}(1990)}]{wangsheeley_90}
Wang, Y.-M. and Sheeley, N. R.~J. (1990), Solar wind speed and coronal
  flux-tube expansion, \emph{Astrophys. J.}, 355, 726

\bibitem[{\textbf{Webb et~al.}(2010)\textbf{Webb, Biesecker, Gopalswamy,
  St.~Cyr, Davila, Thompson et~al.}}]{webb_10b}
Webb, D.~F., Biesecker, D.~A., Gopalswamy, N., St.~Cyr, O.~C., Davila, J.~M.,
  Thompson, B.~J., et~al. (2010), Using stereo-b as an l5 space weather
  pathfinder mission, \emph{Space Research Today}, 178, 10

\bibitem[{\textbf{Webb et~al.}(2018)\textbf{Webb, Gibson, Hewins, McFadden,
  Emery, Malanushenko et~al.}}]{webb_18}
Webb, D.~F., Gibson, S.~E., Hewins, I.~M., McFadden, R.~H., Emery, B.~A.,
  Malanushenko, A., et~al. (2018), Global solar magnetic field evolution over 4
  solar cycles: Use of the mcintosh archive, \emph{Frontiers in Astronomy and
  Space Sciences}

\bibitem[{\textbf{{Wilson}}(1978)}]{wilson_78}
{Wilson}, O.~C. (1978), {Chromospheric variations in main-sequence stars},
  \emph{Astrophys. J.}, 226, 379--396, \doi{10.1086/156618}

\bibitem[{\textbf{{Worden} and {Harvey}}(2000)}]{worden_00}
{Worden}, J. and {Harvey}, J. (2000), {An Evolving Synoptic Magnetic Flux map
  and Implications for the Distribution of Photospheric Magnetic Flux},
  \emph{Solar Phys.}, 195, 247--268, \doi{10.1023/A:1005272502885}

\bibitem[{\textbf{{Wraight} et~al.}(2011)\textbf{{Wraight}, {White}, {Bewsher},
  and {Norton}}}]{wraight_11}
{Wraight}, K.~T., {White}, G.~J., {Bewsher}, D., and {Norton}, A.~J. (2011),
  {STEREO observations of stars and the search for exoplanets}, \emph{Month.
  Not. Roy. Astron. Soc.}, 416, 2477--2493,
  \doi{10.1111/j.1365-2966.2011.18599.x}

\bibitem[{\textbf{{Xiong} and {Liu, Y}}(2014)}]{xiong_14}
{Xiong}, M. and {Liu, Y} (2014), in Taikong,ISSI-BJ Magazine, volume~4,
  volume~4

\bibitem[{\textbf{{Zhao} and {Hoeksema}}(2010)}]{zhaohoek_10}
{Zhao}, X.~P. and {Hoeksema}, J.~T. (2010), {The Magnetic Field at the Inner
  Boundary of the Heliosphere Around Solar Minimum}, \emph{solphys}, 266,
  379--390, \doi{10.1007/s11207-010-9618-0}

\end{thebibliography}

\end{document}